\documentclass[sigplan,screen]{acmart}

\usepackage{booktabs}   %% For formal tables:
\usepackage{caption}
\usepackage[caption=false]{subfig}

\makeatletter\if@ACM@journal\makeatother
\acmJournal{PACMPL}
\acmVolume{1}
\acmNumber{1}
\acmArticle{1}
\acmYear{2017}
\acmMonth{1}
\acmDOI{10.1145/nnnnnnn.nnnnnnn}
\startPage{1}
\else\makeatother
\setcopyright{acmcopyright}
\acmPrice{15.00}
\acmDOI{10.1145/3168812}
\acmYear{2018}
\copyrightyear{2018}
\acmISBN{978-1-4503-5617-6/18/02}
\acmConference[CGO'18]{2018 IEEE/ACM International Symposium on Code Generation and Optimization}{February 24--28, 2018}{Vienna, Austria}
\startPage{1}
\fi

\usepackage{bm}
\usepackage{graphicx}
\usepackage{hhline}
\usepackage{url}
\usepackage{float}
\usepackage{stfloats}
\usepackage{multirow}

\usepackage{listings}

\usepackage[T1]{fontenc}
\usepackage{fancyvrb}

\usepackage[rounded]{syntax}
\usepackage{algorithm}
\usepackage{algorithmicx}
\usepackage{algpseudocode}
\usepackage{flushend}

\newcommand{\ra}[1]{\renewcommand{\arraystretch}{#1}}

\newcommand{\cit}[1]{{\footnotesize\tt #1}}
\newcommand{\eg}{e.g.}
\newcommand{\ie}{i.e.}

\newcommand{\cir}{C-IR}

\newcommand{\R}[0]{\mathbb{R}}

\newcommand{\chol}{{\em potrf}}
\newcommand{\sylv}{{\em trsyl}}
\newcommand{\lyap}{{\em trlya}}
\newcommand{\tinv}{{\em trtri}}
\newcommand{\kf}{{\em kf}}
\newcommand{\gpr}{{\em gpr}}
\newcommand{\lone}{{\em l1a}}

\usepackage{xcolor}
\usepackage{color}

\usepackage{setspace}
\newcommand{\thesys}{\textsc{SLinGen}}
\newcommand{\click}{{\textsc{Cl\makebox[.58\width][c]{1}ck}}}
\newcommand{\clickplain}{{\textsc{Cl{1}ck}}}

\newcommand{\lgen}{\textsc{LGen}}

\definecolor{darkblue}{RGB}{40, 40, 244} 
\definecolor{mydarkgreen}{RGB}{0, 128, 0}
\definecolor{myorange}{rgb}{0.832,0.1914,0}
\definecolor{mygray}{rgb}{0,0,0}
\definecolor{listingbg}{RGB}{240, 240, 240}

\lstdefinestyle{la}{
    language=Matlab,                % choose the language of the code
    basicstyle=\footnotesize\ttfamily,       % the size of the fonts that are used for the code
    numbers=none,                   % where to put the line-numbers
    numberstyle=\footnotesize,      % the size of the fonts that are used for the line-numbers
    stepnumber=1,                   % the step between two line-numbers. If it is 1 each line will be numbered
    showspaces=false,               % show spaces adding particular underscores
    showstringspaces=false,         % underline spaces within strings
    showtabs=false,                 % show tabs within strings adding particular underscores
    tabsize=2,          % sets default tabsize to 2 spaces
    captionpos=b,           % sets the caption-position to bottom
    breaklines=true,        % sets automatic line breaking
    breakatwhitespace=false,    % sets if automatic breaks should only happen at whitespace
    escapeinside={\%*}{*)},          % if you want to add a comment within your code
      aboveskip=1pt,%\smallskipamount,
      belowskip=1pt,%\negsmallskipamount,
    keywordstyle=\bfseries\color{black},
    mathescape=true,
    stringstyle=\color{black},
    moredelim=**[is][\color{gray}]{@}{@},
    commentstyle=\color{gray},
    morekeywords={program, algorithm, operation}
}

\lstdefinestyle{customcpp}{
    language=C++,                % choose the language of the code
    basicstyle=\footnotesize\ttfamily,       % the size of the fonts that are used for the code
    numbers=none,                   % where to put the line-numbers
    numberstyle=\footnotesize,      % the size of the fonts that are used for the line-numbers
    stepnumber=1,                   % the step between two line-numbers. If it is 1 each line will be numbered
    showspaces=false,               % show spaces adding particular underscores
    showstringspaces=false,         % underline spaces within strings
    showtabs=false,                 % show tabs within strings adding particular underscores
    tabsize=2,          % sets default tabsize to 2 spaces
    captionpos=b,           % sets the caption-position to bottom
    breaklines=true,        % sets automatic line breaking
    breakatwhitespace=false,    % sets if automatic breaks should only happen at whitespace
    escapeinside={\%*}{*)},          % if you want to add a comment within your code
      aboveskip=1pt,%\smallskipamount,
      belowskip=1pt,%\negsmallskipamount,
    keywordstyle=\bfseries\color{black},
    mathescape=true,
    stringstyle=\color{black},
    moredelim=**[is][\color{gray}]{@}{@},
    commentstyle=\color{gray}
}

\usepackage{array}
\newcolumntype{I}{!{\vrule width 1.5pt}}
\newlength\savedwidth

\newboolean{IsWide}
\setboolean{IsWide}{true}

\newcommand{\FlaPartition}[2]{
\ifthenelse{\boolean{IsWide}}{{\bf partition } \hspace{-1em} #1 \hspace{-1em} #2}
{{\bf partition } \+ \\ #1 \+ \\ #2 \- \-}
}

\newcommand{\FlaRepartition}[2]{
\ifthenelse{\boolean{IsWide}}{{\bf repartition } \hspace{-1em} #1 \hspace{-1em} #2}
{{\bf repartition } \+ \\ #1 \+ \\ #2 \- \-}
}

\newcommand{\FlaContinue}[1]{
\ifthenelse{\boolean{IsWide}}{{\bf continue with } #1
}
{{\bf continue with } \+ \\ #1 \-
}
}

\newcommand{\blocksize}{1}

\newcommand{\repartitionings}{
\begin{minipage}[t]{3in}
\ \\
\ \\
\ \\
\end{minipage}
}

\newcommand{\repartitionsizes}{ \hspace{ 3.25in} }

\newcommand{\WSrepartition}{
\begin{minipage}[t]{3in}
\ifthenelse{ \equal{\blocksize}{1} }{}
{%
\ifthenelse{ \equal{\blocksize}{blank} }{~}
{{\bf Determine block size $ \blocksize $}} \\
}
{\bf Repartition}
\begin{tabbing}
in \= in \= \+ \kill
\repartitionings \+ \\
{\bf where } \hspace*{-2ex} \repartitionsizes 
\end{tabbing}
\end{minipage}
}

\newcommand{\WSrepartitionNarrow}{
\begin{minipage}[t]{2.05in}
\ifthenelse{ \equal{\blocksize}{1} }{}
{%
\ifthenelse{ \equal{\blocksize}{blank} }{~}
{{\bf Determine block size $ \blocksize $}} \\
}
{\bf Repartition}
\begin{tabbing}
i \= i \= \+ \kill
\repartitionings \+ \\
{\bf where }
\begin{minipage}[t]{1.8in}
\repartitionsizes
\end{minipage}
\end{tabbing}
\end{minipage}
}

\begin{document}

%% Title information
%\title[Short Title]{Full Title}         %% [Short Title] is optional;
\title[Program Generation for Small-Scale Linear Algebra Applications]{Program Generation for Small-Scale \\ Linear Algebra Applications}
%\subtitle{}                     %% \subtitle is optional

                                        %% when present, will be used in
                                        %% header instead of Full Title.
%\titlenote{with title note}             %% \titlenote is optional;
                                        %% can be repeated if necessary;
                                        %% contents suppressed with 'anonymous'
%\subtitle{Subtitle}                     %% \subtitle is optional
%\subtitlenote{with subtitle note}       %% \subtitlenote is optional;
                                        %% can be repeated if necessary;
                                        %% contents suppressed with 'anonymous'

%% Author information
%% Contents and number of authors suppressed with 'anonymous'.
%% Each author should be introduced by \author, followed by
%% \authornote (optional), \orcid (optional), \affiliation, and
%% \email.
%% An author may have multiple affiliations and/or emails; repeat the
%% appropriate command.
%% Many elements are not rendered, but should be provided for metadata
%% extraction tools.

%% Author with single affiliation.
\author{Daniele G. Spampinato }
\affiliation{
  \department{Department of Computer Science}              %% \department is recommended
  \institution{ETH Zurich}            %% \institution is required
  \country{Switzerland}
}
\email{danieles@inf.ethz.ch}          %% \email is recommended

\author{Diego Fabregat-Traver}
\affiliation{
  \department{Aachen Institute for Advanced Study in Computational Engineering Science}              %% \department is recommended
  \institution{RWTH Aachen University}            %% \institution is required
  \country{Germany}
}
\email{fabregat@aices.rwth-aachen.de}

\author{Paolo Bientinesi}
\affiliation{
  \department{Aachen Institute for Advanced Study in Computational Engineering Science}              %% \department is recommended
  \institution{RWTH Aachen University}            %% \institution is required
  \country{Germany}
}
\email{pauldj@aices.rwth-aachen.de}

\author{Markus P\"uschel}
\affiliation{
  \department{Department of Computer Science}              %% \department is recommended
  \institution{ETH Zurich}            %% \institution is required
  \country{Switzerland}
}
\email{pueschel@inf.ethz.ch}

%% Author with two affiliations and emails.
%\author{First2 Last2}
%\authornote{with author2 note}          %% \authornote is optional;
%                                        %% can be repeated if necessary
%\orcid{nnnn-nnnn-nnnn-nnnn}             %% \orcid is optional
%\affiliation{
%  \position{Position2a}
%  \department{Department2a}             %% \department is recommended
%  \institution{Institution2a}           %% \institution is required
%  \streetaddress{Street2a Address2a}
%  \city{City2a}
%  \state{State2a}
%  \postcode{Post-Code2a}
%  \country{Country2a}
%}
%\email{first2.last2@inst2a.com}         %% \email is recommended
%\affiliation{
%  \position{Position2b}
%  \department{Department2b}             %% \department is recommended
%  \institution{Institution2b}           %% \institution is required
%  \streetaddress{Street3b Address2b}
%  \city{City2b}
%  \state{State2b}
%  \postcode{Post-Code2b}
%  \country{Country2b}
%}
%\email{first2.last2@inst2b.org}         %% \email is recommended

%% Paper note
%% The \thanks command may be used to create a "paper note" ---
%% similar to a title note or an author note, but not explicitly
%% associated with a particular element.  It will appear immediately
%% above the permission/copyright statement.
%\thanks{with paper note}                %% \thanks is optional
                                        %% can be repeated if necesary
                                        %% contents suppressed with 'anonymous'

%% Abstract
%% Note: \begin{abstract}...\end{abstract} environment must come
%% before \maketitle command
\begin{abstract}
We present \thesys{}, a program generation system for linear algebra. % The input to \thesys{} is an application mathematically expressed in a linear algebra (LA) language that we define. LA provides basic matrix/vector/scalar operations, higher level operations including triangular solve, Cholesky and LU factorization, and loops.
The input to \thesys{} is an application expressed mathematically in a linear-algebra-inspired language (LA) that we define. LA provides basic scalar/vector/matrix additions/multiplications and higher level operations including linear systems solvers, Cholesky and LU factorizations.
The output of \thesys{} is performance-optimized single-source C code, optionally vectorized with intrinsics. The target of \thesys{} are small-scale computations on fixed-size operands, for which a straightforward implementation using optimized libraries (e.g., BLAS or LAPACK) is known to yield suboptimal performance (besides increasing code size and introducing dependencies), but which are crucial in control, signal processing, computer vision, and other domains. Internally, \thesys{} uses synthesis and DSL-based techniques to optimize at a high level of abstraction.
We benchmark our program generator on three prototypical applications: the Kalman filter, Gaussian process regression, and an L1-analysis convex solver, as well as basic routines including Cholesky factorization and solvers for the continuous-time Lyapunov and Sylvester equations. The results show significant speed-ups compared to straightforward C with Intel icc and clang with a polyhedral optimizer, as well as library-based and template-based implementations.
\end{abstract}

%\ccsdesc[100]{Software and its engineering~Specialized application languages}
%\ccsdesc[100]{Computing methodologies~Symbolic and algebraic manipulation}
%\ccsdesc[100]{Computing methodologies~Linear algebra algorithms}
%% End of generated code

%% Keywords
%% comma separated list
%\keywords{keyword1, keyword2, keyword3}  %% \keywords is optional

%% \maketitle
%% Note: \maketitle command must come after title commands, author
%% commands, abstract environment, Computing Classification System
%% environment and commands, and keywords command.
\maketitle
\renewcommand{\shortauthors}{D. G. Spampinato, D. Fabregat-Traver, P. Bientinesi, and M. P\"{u}schel}

\section{Introduction}\label{sec:intro}

A significant part of processor time worldwide is spent on mathematical algorithms that are used in simulations, machine learning, control, communication, signal processing, graphics, computer vision, and other domains. The problem sizes and computers used range from the very large (e.g., simulations on a supercomputer or learning in the cloud) to the very small (e.g., a Kalman filter or Viterbi on an embedded processor). Both scenarios have in common the need for very fast code, for example to save energy, to enable real-time, or to maximize processing resolution. The mathematics used in these domains may differ widely, but the actual computations in the end often fall into the domain of linear algebra, meaning sequences of computations on matrices and vectors. 

For large-scale linear algebra applications, the bottleneck is usually in cubic-cost components such as matrix multiplication, 
matrix decompositions, and solving linear systems. High performance is thus attained by using existing highly optimized libraries (typically built around the interfaces of BLAS~\cite{Dongarra:90} and LAPACK~\cite{Anderson:99}). For small-scale applications, the same libraries are not as optimized, may incur overhead due to fixed interfaces and large code size, and introduce dependencies~\cite{Spampinato:14}. The small scale is the focus of this paper.

\paragraph{Program generation for small-scale linear algebra.} In an ideal world, a
programmer would express a linear algebra computation as presented in a book
from the application domain, and the computer would produce code that is
both specialized to this computation and highly optimized for the target
processor. With this paper, we make progress towards this goal and introduce
\thesys{}, a generator for small-scale linear algebra applications on fixed-size operands.
The input to \thesys{} is the application expressed in a linear algebra language (LA) that we introduce. LA provides basic scalar/vector/matrix operations, higher level operations including triangular solve, Cholesky and LU factorization, and loops. The output is a single-threaded, single-source C code, optionally vectorized with intrinsics.

\begin{table}
\footnotesize
\caption{Kalman filter (one iteration at time step $k$). Matrices are upper case, vectors lower
  case. During prediction (steps \eqref{eq:prx} and \eqref{eq:prp}) a new estimate of the system state $x$ is computed \eqref{eq:prx} along with an associated covariance matrix $P$. During update (steps \eqref{eq:updx} and \eqref{eq:updp}), the predictions are combined with the current observation $z$.}\label{tab:kalman}
\begin{align}
    x_{k\mid k-1} = &\ F x_{k-1\mid k-1} + B u \label{eq:prx} \\
%    \smash{\raisebox{\dimexpr.5\normalbaselineskip+.5\jot}{$%
%            \text{Predict: }\left\{\begin{array}{@{}c@{}}\\[\jot]\\[\jot]\end{array}\right.$}}
    P_{k\mid k-1} = &\ F P_{k-1 \mid k-1} F^T + Q \label{eq:prp}\\
    x_{k\mid k}   = &\ x_{k\mid k-1} + P_{k\mid k-1} H^T \notag \\
                    & \times(H P_{k\mid k-1} H^T + R)^{-1}( z_k - H x_{k\mid k-1} ) \label{eq:updx} \\
    P_{k\mid k}   = &\ P_{k\mid k-1} - P_{k\mid k-1} H^T \notag \\
%    \smash{\raisebox{\dimexpr 1.5\normalbaselineskip+1.5\jot}{
%    $\text{Update: }\left\{\begin{array}{@{}c@{}}\\[\jot]\\[\jot]\\[\jot]\\[\jot]\\[\jot]\end{array}\right.$}}
                    \phantom{P_{k\mid k-1} = }& \times(H P_{k\mid k-1} H^T + R)^{-1}H P_{k\mid k-1}\; \label{eq:updp}
\end{align}
\end{table}

As illustrating example, we consider the Kalman filter, which is ubiquitously
used to control dynamic systems such as vehicles and robots~\cite{Scharf:91},
and a possible input to \thesys{}. Table~\ref{tab:kalman} shows a basic Kalman
filter, which performs computations on matrices (upper case) and vectors
(lower case) to iteratively update a (symmetric) covariance matrix
$P$. The filter performs basic multiplications and additions on matrices and vectors, but also requires a Cholesky factorization and a triangular solve to perform the matrix inversions. Note that operand sizes are typically fixed (number of states and measurements in the system) and in the 10s or even smaller. We show benchmarks with \thesys{}-generated code for the Kalman filter later.

\begin{figure}
    \centering
    \includegraphics[width=.9\columnwidth]{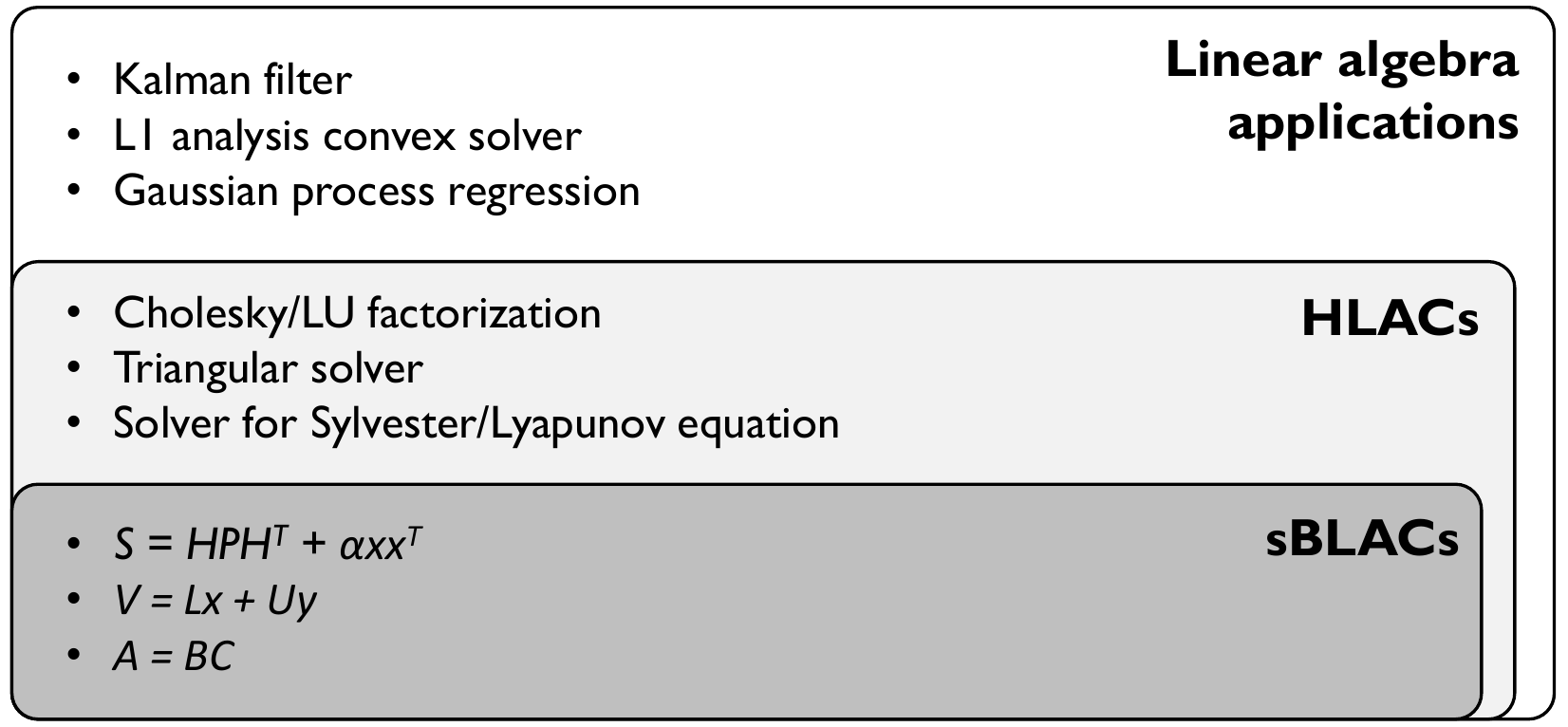}
    \caption{Classes of linear algebra computations used in this paper with examples.
%Basic linear algebra computations possibly with structured matrices (sBLACs), higher-level computations (HLACs), and entire linear algebra applications, such as the Kalman filter.
}
    \label{fig:compclasses}
\end{figure}

\paragraph{Classification of linear algebra computations.}
Useful for this paper, and for specifying our contribution, is an organization of the types of computations that our generator needs to process to produce single-source code. We consider the following three categories (Fig.~\ref{fig:compclasses}):
\begin{itemize}
\item \emph{Basic linear algebra computations, possibly with structured matrices (sBLACs, following~\cite{Spampinato:14}):} Computations on matrices, vectors, and scalars using basic operators: multiplication, addition, and transposition. Mathematically, sBLACs include (as a subset) most of the computations supported by the BLAS interface. 
    \sloppypar
\item \emph{Higher-level linear algebra computations (HLACs):} Cholesky and LU
  factorization, triangular solve, and other direct solvers. 
Mathematically, HLAC algorithms (in particular when blocked for performance) are expressed as loops over sBLACs.
\item \emph{Linear algebra applications:}  Finally, to express and support entire applications like the Kalman filter, this class includes loops and sequences of sBLACs, HLACs, and auxiliary scalar computations. 
\end{itemize}

\thesys{} supports the latter class (and thus also HLACs), and thus can generate code for a significant class of real-world linear algebra applications. In this paper we consider three case studies: the Kalman filter, Gaussian process regression, and L1-analysis convex optimization.

\paragraph{Contributions.}
In this paper, we make the following main contributions.
\begin{itemize}
\item The design and implementation of a domain-specific system that generates performant, single-source C code directly from a high-level linear algebra description of an application. This includes the definition of the input language LA, the use of synthesis and DSL (domain-specific language) techniques to optimize at a high, mathematical level of abstraction, and the ability to support vector ISAs. The generated code is single-threaded since the focus is on small-scale computations.
\item As a crucial component, we present the first generator for a class of HLACs that produces single-source C (with intrinsics), i.e., does not rely on a BLAS implementation.
\item Benchmarks of our generated code for both single HLACs and application-level linear algebra programs comparing against hand-written code: straightforward C optimized with Intel icc and clang with the polyhedral Polly, the template-based Eigen library, and code using libraries (Intel's MKL, ReLAPACK, RECSY).
\end{itemize}
As we will explain, part of our work builds on, but considerably expands, two prior tools: the sBLAC compiler \lgen{} and \click{}.
\lgen{}~\cite{Spampinato:14,Spampinato:16} generates vectorized C code for single sBLACs with fixed operand sizes; in this paper we move considerably beyond in functionality to generate code for an entire linear algebra language that we define and that includes HLAC algorithms (which themselves involve loops over sequences of sBLACs, some with changing operand size) and entire applications. \click{}~\cite{cl1ck-pme,cl1ck-linv} synthesizes blocked algorithms (but not C code) for HLACs, represented in a DSL assuming an available BLAS library as API. Our goal is to generate single-source, vectorized C code for HLACs and entire linear algebra applications containing them.

\section{Background}\label{sec:bg}

We provide background on the two prior tools that we use in our work. \lgen{}~\cite{Spampinato:14,Kyrtatas:15,Spampinato:16} compiles single sBLACs (see Fig.~\ref{fig:compclasses}) on fixed-size operands into (optionally vectorized) C code. \click{}~\cite{cl1ck-pme,cl1ck-linv} generates blocked algorithms for (a class of) HLACs, expressed at a high level using the BLAS API. 

In the following, we use uppercase, lowercase, and Greek letters to denote matrices, vectors, and scalars, respectively.

\subsection{\lgen{}}\label{sec:lgen}

An sBLAC is a computation on scalars, vectors, and (possibly structured) matrices that involves addition, multiplication, scalar multiplication, and transposition. Examples include $A = BC$, $y = Lx + Uy$, $S = HPH^T + \alpha xx^T$, where $S$ is symmetric and $L$/$U$ are lower/upper triangular. The output is on the left-hand side and may also appear as input.

\begin{figure}
  \centering
  \includegraphics[scale=0.4]{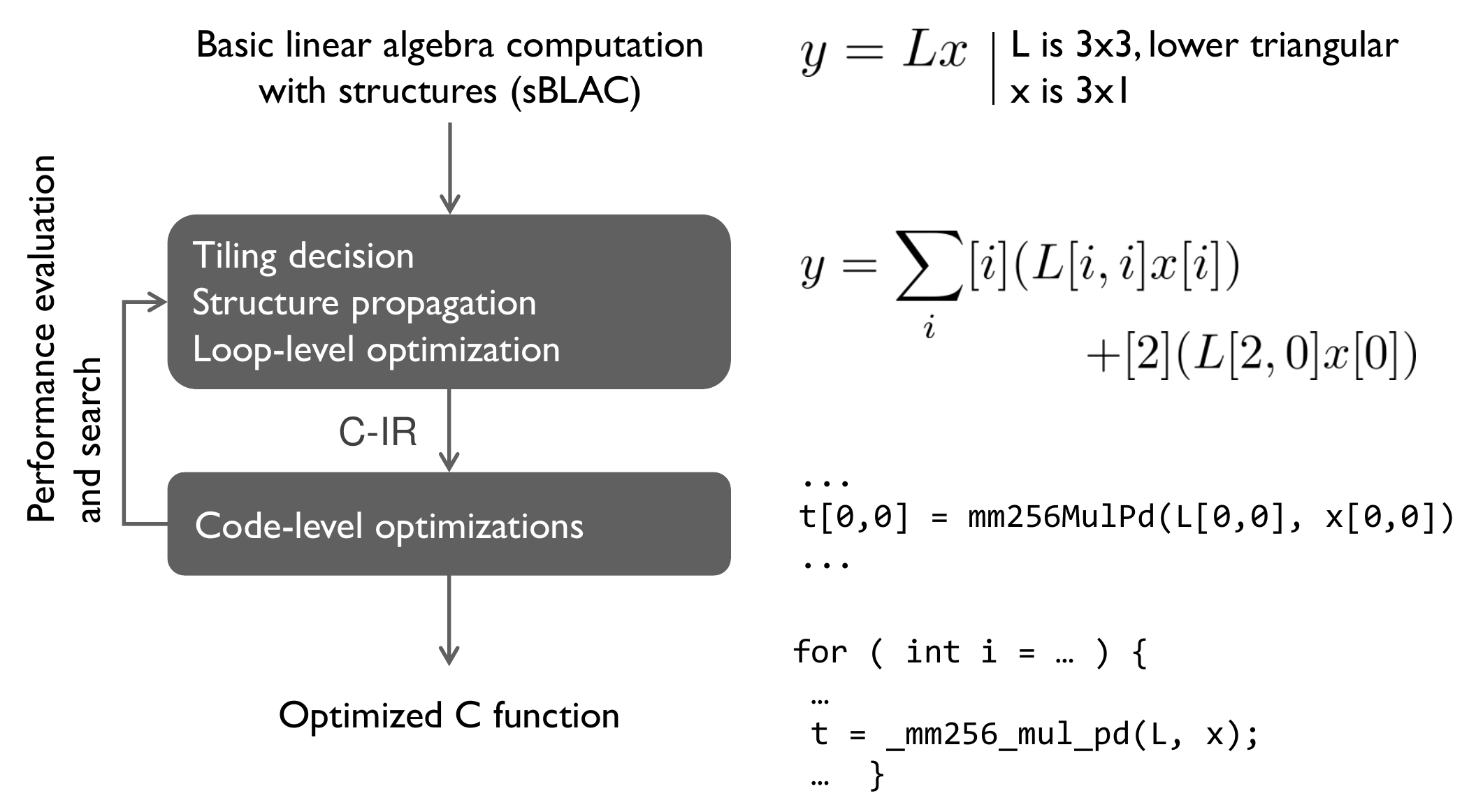}
  \caption{\label{fig:lgen}The architecture of \lgen{} including a sketched small example.}
%  On the side example processing of a small matrix-vector multiplication between a $4\times 4$ triangular matrix $L$ and a vector $x$. The sBLAC is tiled with tile size 2. During the first phase tiling and structure propagation are used to generate a loop-based formulation of the sBLAC where triangular matrix-vector multiplications on diagonal are split from the general multiplication at the bottom-left corner. Here $[i,i]$ indicates the interval access $(i:i+2, i:i+2)$. If the access appears on the right of an expression it shows at what interval of the output the result of its computation should be stored.}
\end{figure}

\paragraph{Overview.}
\lgen{} translates an input sBLAC on fixed-size operands into C code using two intermediate compilation phases as shown in Fig.~\ref{fig:lgen}. During the first phase, the input sBLAC is optimized at the mathematical level, using a DSL that takes possible matrix structures into account. These optimizations include multi-level tiling, loop merging and exchange, and matrix structure propagation. During the second phase, the mathematical expression obtained from the first phase is translated into a C-intermediate representation. At this level, \lgen{} performs additional optimizations such as loop unrolling and scalar replacement. Finally, since different tiling decisions lead to different code versions of the same computation, \lgen{} uses autotuning to select the fastest version for the target computer.

\paragraph{Explicit vectorization.}
\lgen{} allows for multiple levels of tiling and arbitrary tile sizes. If vectorization is enabled, the innermost level of tiling 
decomposes an expression into so-called $\nu$-BLACs, where $\nu$ is the vector length (e.g., $\nu=4$ for double precision AVX). There are 18 $\nu$-BLACs (all single operations on $\nu\times\nu$ matrices and vectors of length $\nu$), which are pre-implemented once for a given vector ISA together with vectorized data access building blocks, called Loaders and Storers, that handle leftovers and structured matrices.

\subsection{\clickplain{}}\label{sec:cl1ck}

\click{}~\cite{cl1ck-pme,cl1ck-linv} is an algorithm generator that implements the FLAME methodology~\cite{Bientinesi:05}.
The input to the generator is an HLAC expressed in terms of the standard matrix
operators---addition, product, transposition and inversion---and  matrix properties such as orthogonal, symmetric positive definite, and triangular. The output is a family of loop-based algorithms which make use of existing
BLAS kernels; while all the algorithms in a family are mathematically equivalent, they provide a range
of alternatives in terms of numerical accuracy and performance. 

\begin{figure}
  \centering
  \includegraphics[width=0.9\columnwidth]{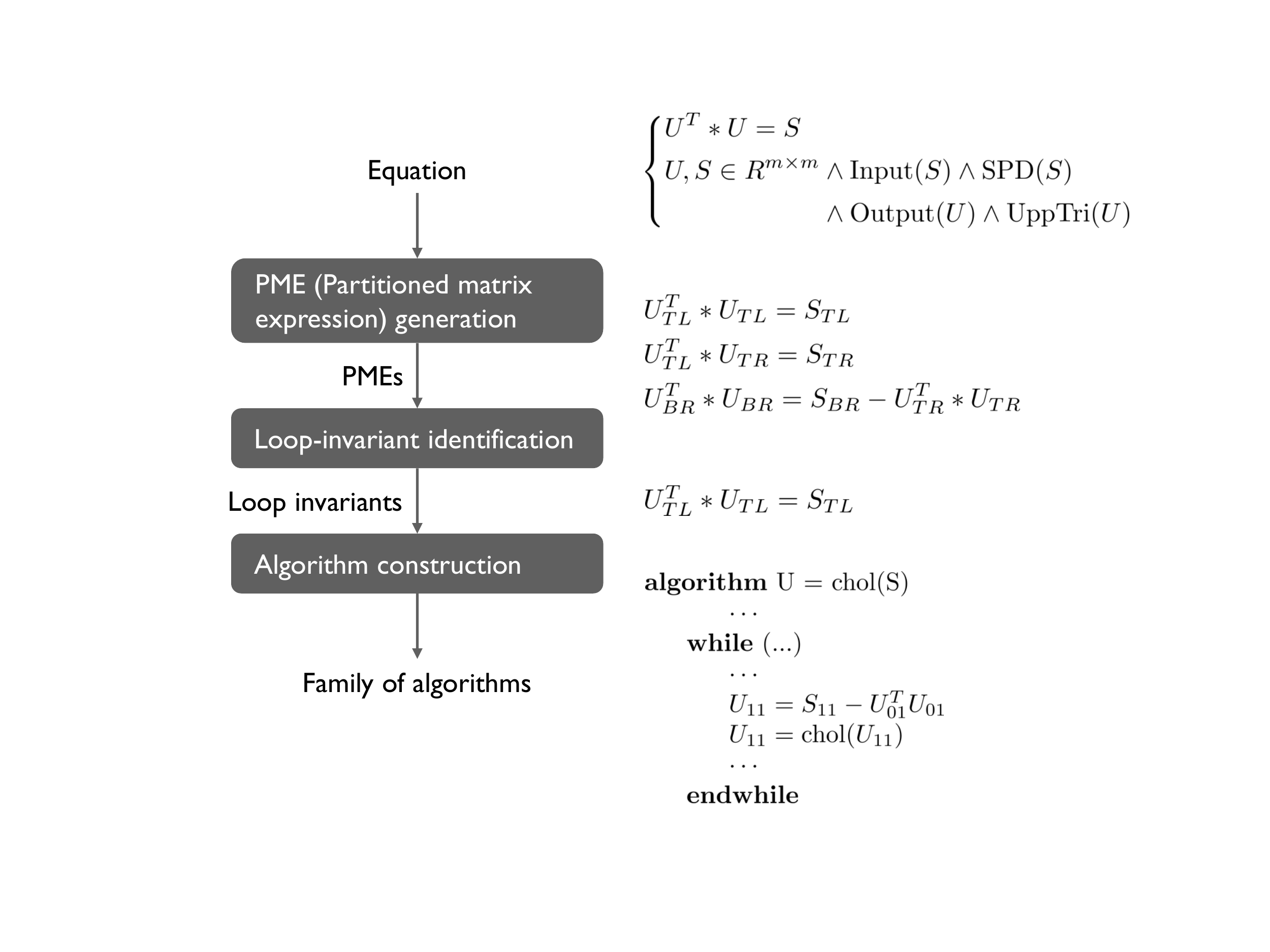}
  \caption{{\sc Cl1ck}: For a target input computation, a family of algorithms
    is generated in three stages. 
  }
  \label{fig:click}
\end{figure}

\paragraph{Overview.}
As illustrated in Fig.~\ref{fig:click}, the generation takes place in three
stages: PME Generation, Loop Invariant Identification, and Algorithm
Construction.
In the first stage, the input equation (e.g., $U^T*U = S$, where $U$ is the output), is blocked
symbolically to obtain one or more recursive formulations, called ``Partitioned Matrix
Expression(s)'' (PMEs); this stage makes use of a linear algebra knowledge-base,
as well as pattern matching and term rewriting. In our example, blocking yields\footnote{
    $T$, $B$, $L$, and $R$ stand for Top, Bottom, Left, and Right, respectively.}% (TL = top left, etc.)
$$
\left(\begin{smallmatrix}U_{TL}^T & 0 \\ U_{TR}^T & U_{BR}^T\end{smallmatrix}\right)
\left(\begin{smallmatrix}U_{TL} & U_{TR}\\ 0 & U_{BR}\end{smallmatrix}\right) =
\left(\begin{smallmatrix}S_{TL} & S_{TR}\\ S_{BL} & S_{BR}\end{smallmatrix}\right),
$$
from which the three dependent equations shown are generated.
In the second stage, \click{} identifies possible loop invariants for the yet-to-be-constructed
loop-based algorithms. One example here is the predicate $U_{TL}^T U_{TL} = S_{TL}$ that has to be satisfied at the
beginning and the end of each loop iteration. Finally, each loop invariant is translated into an algorithm 
that expresses the computation in terms of recursive calls, and BLAS- and LAPACK-compatible
operations. As the process completes, the input equation becomes ``known'': it is assigned
a name, is added to the knowledge base, and can be used as a building block for more complex operations.

\paragraph{Formal correctness.}
The main idea behind \click{} and the underlying FLAME methodology is
the simultaneous construction of a loop-based blocked algorithm and its proof
of correctness. This is accomplished by identifying the loop invariant that the
algorithm will have to satisfy, {\em before} the algorithm exists. From the
invariant, a template of a proof of correctness is derived, and the algorithm is
then constructed to satisfy the proof.

\begin{figure}
%{
%    \footnotesize
%    \renewcommand{\theFancyVerbLine}{\rmfamily\tiny\arabic{FancyVerbLine}:}
%%\begin{Verbatim}[frame=single,commandchars=\\\{\},codes={\catcode`$=3\catcode`^=7},numbers=left,
%\begin{Verbatim}[commandchars=\\\{\},codes={\catcode`$=3\catcode`^=7},numbers=left,
%    numberblanklines=false,numbersep=-8pt]
%   la-program  := \{declaration\} \{equation\}
%   declaration := type id '<' [ size [ ',' size ] ',' ]
%                   iotype \{',' property\} [',' ow] '>' ';';
%   iotype      := 'In' | 'Out';
%   property    := 'LoTri' | 'UpTri' | 'UpSym' | 'LoSym' 
%                | 'PD' | 'NS' | 'UnitDiag';
%   ow          := 'ow(' id ')';
%   statement    := sBLAC | HLAC ';';
%   sBLAC    := id '=' expression ;
%   HLAC    :=  expression '=' expression | 
%                      id '=' 'inv(' id ')' ;
%
%\end{Verbatim}
%}
\setlength{\grammarindent}{7em}
\begin{grammar}\small
<la-program> ::= \{<declaration>\} \{<statement>\}

<declaration> ::= `Mat' <id> `(' <size> `,' <size> `)'
               `<' <iotype> \{ `,' <property> \} [ `,' <ow>] `>;' 
               \alt `Vec' <id> ... | `Sca' <id> ...

   <iotype> ::= `In' | `Out' | `InOut'
   
   <property> ::= `LoTri' | `UpTri' | `UpSym' | `LoSym' 
                \alt `PD' | `NS' | `UnitDiag'
                
   <ow> ::= `ow(' <id> `)'
   
   <statement> ::= <for-loop> | <sBLAC> | <HLAC> `;'

   <for-loop> ::= `for (i = ...) \{' \{<statement>$_i$ \} `\}'
   
   <sBLAC> ::= <id> `=' <expression> 
   
   <HLAC> ::=  <expression> `=' <expression> 
   		\alt <id> `=' `(' <id> \lit{$)^{-1}$}

\end{grammar}
\caption{Grammar for the LA language. An LA program consists of the declaration
    of a number of operands, and a sequence of computational statements.
    Operands may be scalars, vectors, or matrices, which are declared as either
    input (\protect\lit{In}) or output (\protect\lit{Out}) and of a certain
    size.
    The non-terminals \synt{id} and \synt{size} are any variable name and fixed
    size integer respectively.
    Matrices can have one or more properties. Structural properties beginning
    with \protect\lit{Up} and \protect\lit{Lo} specify respectively upper and
    lower storage format for both triangular and symmetric matrices.
    \protect\lit{PD} and \protect\lit{NS} stand for positive definiteness and
    non-singularity respectively.  \synt{expression} represents any
    well-defined combination of input and output scalars, vectors and matrices
    with operators \protect\lit{+}, \protect\lit{-}, \protect\lit{*},
    \protect\lit{$(\cdot)^\text{T}$} (transposition). If an expression is
    defined exclusively over scalars then it can include also the division
    (\protect\lit{/}) and square root (\protect\lit{$\sqrt{}$}) operators.
    When an \synt{id} appears alone on the left-hand side, it must be an output
    element and it can also appear on the right-hand side
    (\protect\lit{InOut}).
    The notation \synt{statement}$_i$ indicates that
    data accesses in a statement might depend on the induction variable of the surrounding loop.}
    \label{fig:lagrammar}
\end{figure}

\subsection{Challenges in connecting \clickplain{} and \lgen{}}

In this paper we address two main research challenges. The first one is how to connect \click{} and \lgen{} to generate code for HLACs. 
%This requires an extension from single sBLACs (the domain of \lgen{}) to the entire DSL used by \click{} to express its outputs (see Fig.~\ref{fig:click}), including multiple statements, while loops, changing block sizes, and HLACs on small blocks. 
This problem requires 1) an extension from single sBLACs (the domain of \lgen{}) to the entire DSL used by \click{} to express its loop-based outputs with multiple statements (see Fig.~\ref{fig:click}), and 2) a way to automatically synthesize HLACs on small blocks (Sec.~\ref{sec:crossfuncopts}). 
%Further, just inlining generated C code for the BLAS calls (interpreted as sBLACs) in \click{}'s output misses important optimizations that can be done across sBLACs for better vectorization (Sec.~\ref{sec:vec}) and locality (Sec.~\ref{sec:locality}). 
The second challenge consists is designing a code generation approach that extends across all statements, both at the level of a single HLAC and for an entire linear algebra program.
%Generating and inlining C code for each statement in \click{}'s output misses important optimizations that can lead to better vectorization (Sec.~\ref{sec:vec}) and locality (Sec.~\ref{sec:locality}). 
When the generation and inlining of C code is performed independently for each statement in \click{}'s output, certain optimizations cannot be applied, thus missing important opportunities for better vectorization (Sec.~\ref{sec:vec}) and locality (Sec.~\ref{sec:locality})
% To demonstrate these challenges, and the efficacy of our solution, we provide later detailed benchmarks with generated HLAC code, including comparisons to straightforward C and the prior \click{} code that calls optimized BLAS functions.
In Sec.~\ref{sec:results}, we demonstrate the efficacy of our solution compared to previous work on selected benchmarks. For a set of HLAC benchmarks (as previously mentioned, \click{} only takes HLACs as an input), we include the implemention of \click{}'s generated algorithms using optimized BLAS and LAPACK functions among our competitors. 

%A connection between the two levels at which \click{} and \lgen{} operate is important to achieve our goal but not enough. A simple invocation of \lgen{} on \click{}'s output would miss out on important steps that we found fundamental to obtain the performance results shown in this paper. Before our work, \click{} generated high-level functions that call BLAS and LAPACK routines, \lgen{} only code for a single-statement basic linear algebra computation. The generation of highly-optimized code for entire linear algebra applications using a complete and automated push-button methodology requires the support of an entire language as discussed next in Sec.~\ref{sec:slingen}. In particular, this includes the ability to optimize across the occurring sBLACs in the given application for better vectorization (Sec.~\ref{sec:vec}) and locality (Sec.~\ref{sec:locality}). In Sec.~\ref{sec:results}, we compare our approach with \click{}'s for a set of HLAC benchmarks where \click{}'s generated algorithms are implemented using functions from a high-performance linear algebra library. 
 
\section{\thesys{}}\label{sec:slingen}

We present \thesys{}, a program generator for small-scale linear algebra applications. The input to \thesys{} is a linear algebra program written in LA, a MATLAB-like domain-specific language described by the grammar in Fig.~\ref{fig:lagrammar}.  The output is a single-source optimized C function that implements the input, optionally vectorized with intrinsics.

LA programs are composed of sBLACs and HLACs over scalars, vectors, and matrices of a fixed size.  As an example, the program in Fig.~\ref{fig:laprog} shows a slightly modified fragment of the Kalman filter in Table~\ref{tab:kalman}.
In this program, statement 7 corresponds to an sBLAC that computes a matrix multiplication and an addition of symmetric matrices. Statements 8 and 9 are two HLACs: the Cholesky decomposition of $S$ and a triangular linear system with unknown $B$. Note that, without an explicit input and output specification, it would be impossible to tell the difference between 8 and 9.

%      Vec $u(m), z(k)$ <Input>;
%      Vec $v_0(k), x(n), y(n)$ <InOut>;
%      Vec $v_1(k)$ <In,  overwrites($vx_0$)>;
%
%      Mat $B(n,m), F(n,n), H(k,n)$ <In>;
%      Mat $M_0(n,n),M_1(k,n),M_2(n,k)$ <InOut>;   
%      Mat $M_3(k,k), P(n,n), Y(n,n)$ <InOut, SPD, Up>;
%      Mat $Q(n,n), R(k,k)$ <In, SPD, Up>;
%      Mat $U_i(k,k)$  <In, UpTri, NS, ow($M_3$)>;
%      Mat $U_o(k,k)$  <Out, UpTri, NS, ow($M_3$)>;
%      Mat $M_4(k,n)$ <Out, ow($M_1$)>;
%    
%      $y = F*x + B*u$;
%      $Y = F*P*F^T + Q$;
%      $v_0 = z - H*y$;
%      $M_1 = H*Y$;    
%      $M_2 = Y*H^T$;
%      $M_3 = M_1*H^T + R$;
%      $U_o^T*U_o   = M_3$;
%      $U_i^T*v_1 = v_0$;
%      $U_i*v_1 = v_0$;
%      $U_i^T*M_4 = M_1$;
%      $U_i*M_4 = M_1$;
%      $x = y + M_2*v_0$;
%      $P = Y - M_2*M_1$;
\begin{figure}

\begin{lstlisting}[style=la, numbers=left, numbersep=-8pt, numberstyle=\tiny\color{gray}]
  Mat $H(k,n)$ <In>;
  Mat $P(k,k), R(k,k)$ <In, UpSym, PD>;
  Mat $S(k,k)$ <Out, UpSym, PD>;
  Mat $U(k,k)$ <Out, UpTri, NS, ow($S$)>;
  Mat $B(k,k)$ <Out>;

  $S = H*H^T + R$;
  $U^T * U = S$;
  $U^T * B = P$;
\end{lstlisting}
\caption{Example LA program for given constants $n$ and $k$. Note the explicit specification of input and output matrices.}
\label{fig:laprog}
\end{figure}

\begin{figure}
    \includegraphics[width=0.45\textwidth]{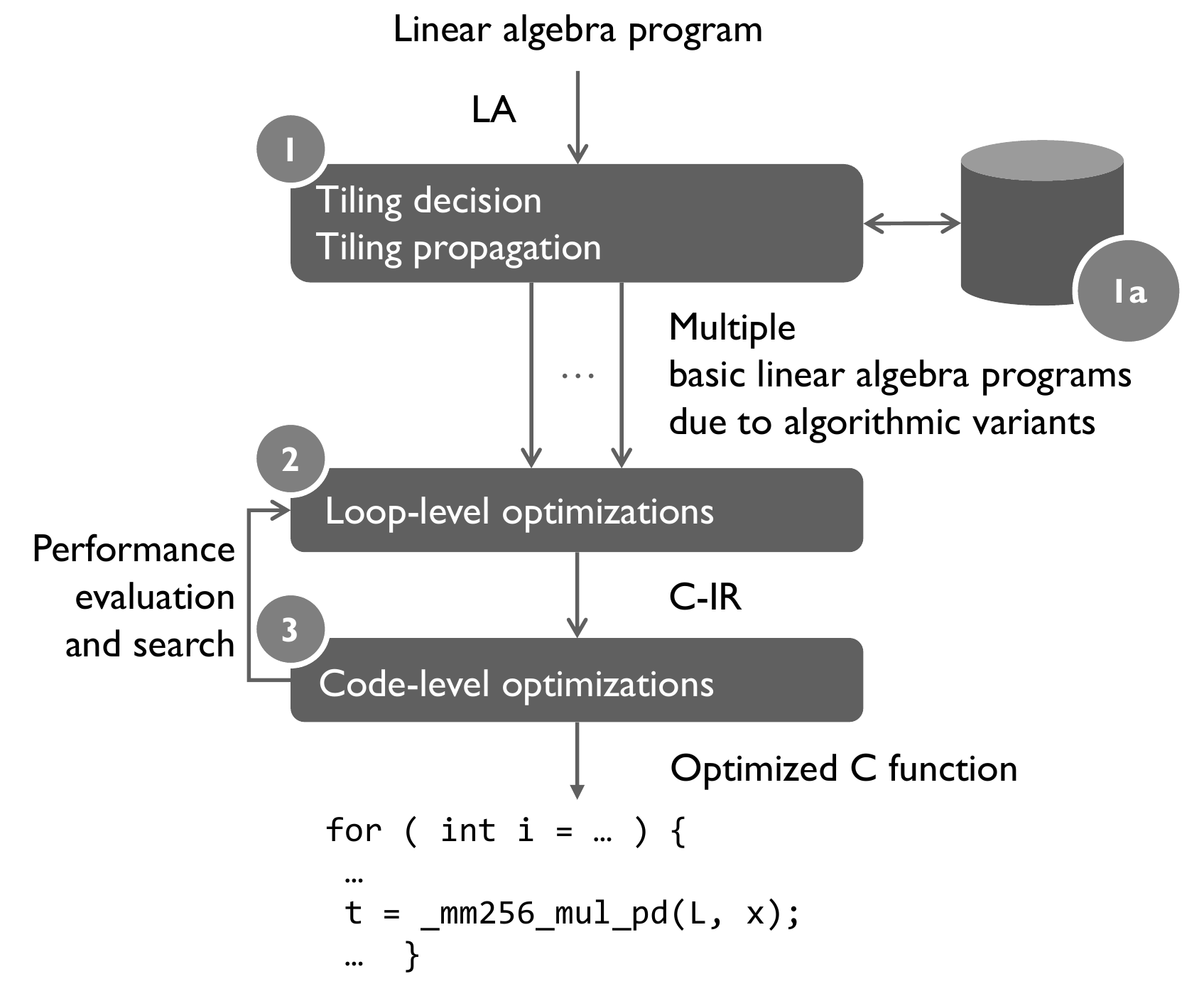}
    \caption{Architecture of \thesys{}.}
    \label{fig:arch}
\end{figure}

\paragraph{Overview.} 
The general workflow of \thesys{} is depicted in Fig.~\ref{fig:arch}. Given an
input LA program, the idea is to perform a number of lowering steps, until the
program is expressed in terms of pre-implemented vectorized codelets and
auxiliary scalar operations (divisions, square roots, ...).  In the first stage,
the input program is transformed into one or more alternative {\em basic LA programs}---implementations that rely exclusively on sBLACs and scalar operations. To achieve this, loop-based algorithms are synthesized for each occurring HLAC. 
Each of these expands the HLAC into a computation on sBLACs and scalars. 
Since more than one algorithm is available for each HLAC, autotuning can be
used to explore different alternatives.

Next, each basic LA program is processed in two additional stages, as shown in
Fig.~\ref{fig:arch}. In Stage 2, each implementation is translated into a C-like intermediate representation (C-IR); if vectorization is enabled, rewriting
rules are used to increase vectorization opportunities.
For example, a group of
scalar statements can be rewritten as a vectorizable sBLAC. C-IR includes (1) special pointers for accessing portions of matrices and vectors, (2) mathematical operations on the latter, and (3) \texttt{For} and \texttt{If} constructs with affine conditions on induction variables. In Stage 3 of
Fig.~\ref{fig:arch}, each C-IR program is unparsed into C code and its
performance is measured. During the translation, code-level
optimizations are applied. As an example, if vectorization is desired, a domain-specific
load/store analysis is applied to replace explicit memory loads and stores with
more efficient data rearrangements between vector variables.  
Finally, the code with best performance is selected as the final output.

The three stages are now explained in greater detail.

\subsection{Stage 1: Basic Linear Algebra Programs Synthesis}\label{sec:crossfuncopts}

In Stage 1, \thesys{} identifies all HLACs in the input program and synthesizes for each of them algorithms that consist of {\em basic building blocks}, i.e., sBLACs and scalar operations.

\paragraph{Identifying HLACs.}
\thesys{} traverses all statements in the input LA program collecting all
HLACs. As described by the LA grammar in Fig.~\ref{fig:lagrammar}, HLACs are
characterized by the presence of an expression on their left-hand side, or by
the use of a non-basic operator on the right-hand side (in our case we only
consider the inverse). As an example, given the LA input in
Fig.~\ref{fig:laprog}, \thesys{} would collect the two HLACs on lines 8 and 9.

\paragraph{Translation into basic form.}
Next, \thesys{} derives for each HLAC a number of loop-based algorithms.  This process is implemented using an iterative
extension of the \click{} methodology (see Sec.~\ref{sec:cl1ck}), which we describe using as running example line~8 of Fig.~\ref{fig:laprog}:
\begin{equation}\label{eq:runex}
U^TU = S.
\end{equation}
Both matrices are of general but fixed size $m \times m$.
As we explain next, the translation first decomposes the HLAC into operations on sBLACs and small HLACs of vector-size $\nu$, and then constructs codelets that consist of small sBLACs and scalar operations.
  
\paragraph{1) Refinement of HLACs.}
The refinement partitions the HLAC along each of its dimensions until only sBLACs and vector-size HLACs remain. Therefore, the first decision is how to partition the dimensions. 
For the example in~\eqref{eq:runex}, \thesys{} decides to partition along rows and columns of both
matrices to carry on and exploit the symmetry of $S$ and the triangular structure of $U$.
While any number of levels of partitioning are possible, for the sake of
brevity, we assume that only two levels of partitioning are desired, with blockings of
size $\nu$ (the architecture vector width) and 1.
With the chosen partitioning, three algorithms are obtained for~\eqref{eq:runex} (associated with three possible loop invariants);
we continue with the one shown in Fig.~\ref{fig:cholalgo}. This algorithm relies on 
two matrix multiplications (sBLACs on lines 2 and 4), one HLAC of size $\nu \times m$ (\ie{}, the linear system on line 5), and one HLAC of vector-size (\ie{}, the recursive instance of~\eqref{eq:runex} on line 3).
\newsavebox{\lstlafrag}
\begin{lrbox}{\lstlafrag}
\begin{minipage}{.5\columnwidth}
\begin{lstlisting}[style=la, numbers=left, numbersep=-8pt, numberstyle=\tiny\color{gray}]
  for(i=0; i<m; i+=$\nu$) {
      $S_{BL} = S_{BL} - S_{TL}^T * S_{TL} $;
      $\underline{U}_{BL}^T * \underline{U}_{BL} = S_{BL}$;
      $S_{BR} = S_{BR} - S_{TL}^T S_{TR}$;
      $U_{BL} * \underline{U}_{BR} = S_{BR}$;
  }
\end{lstlisting}
\end{minipage}
\end{lrbox}

\newsavebox{\matpart}
\begin{lrbox}{\matpart}
\begin{minipage}{.5\columnwidth}
\includegraphics[width=.8\columnwidth]{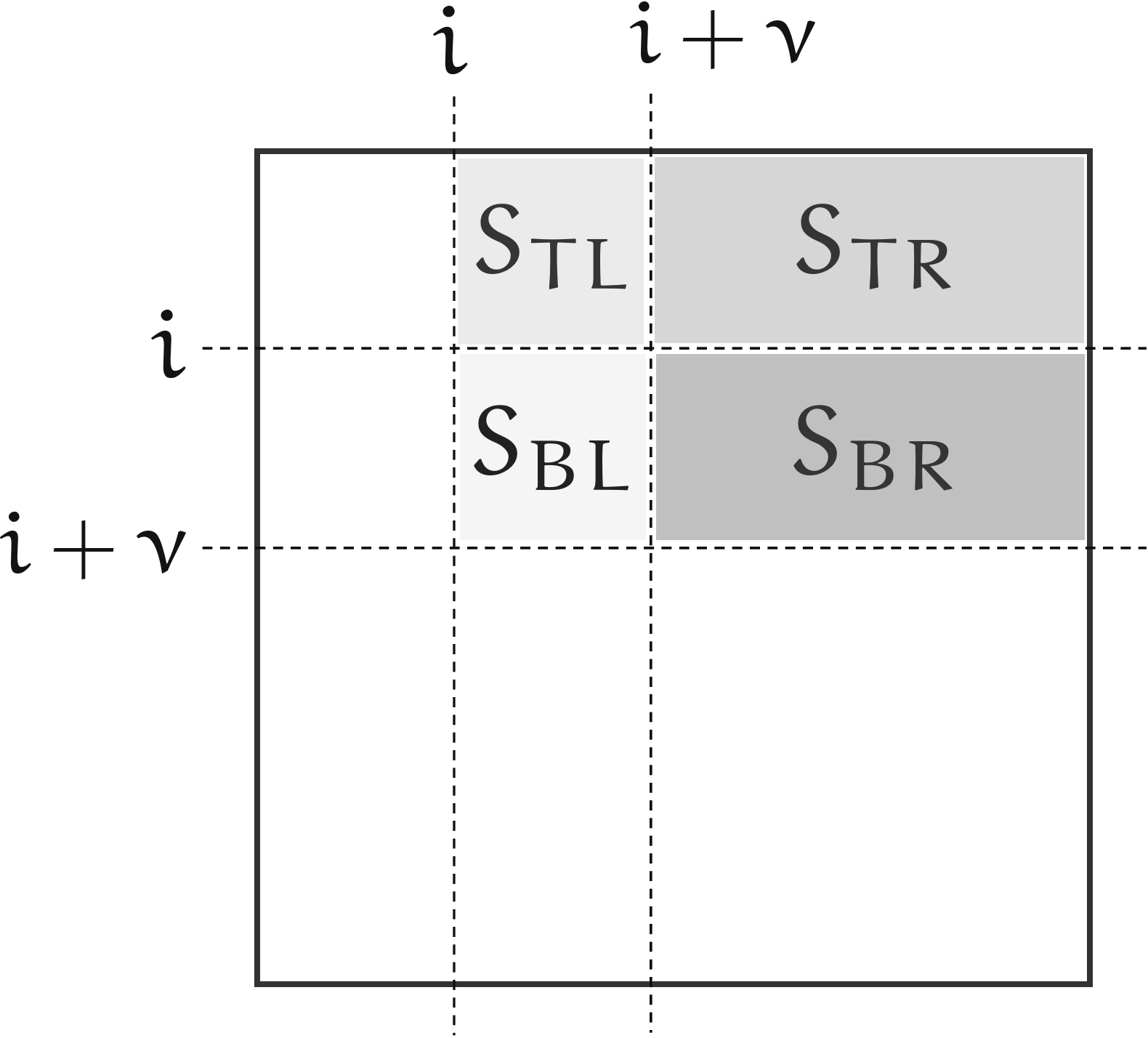} 
\end{minipage}
\end{lrbox}

\begin{figure}
\subfloat[]{\usebox\lstlafrag \label{subfig:lafrag}}
\subfloat[]{\usebox\matpart \label{subfig:matpart}}
\caption{\protect\subref{subfig:lafrag} One possible synthesized LA fragment for the HLAC in~\eqref{eq:runex}. \protect\subref{subfig:matpart} shows partitions of $S$ involved in the computation; $U$ is partitioned analogously. 
%Matrix $M$ is a temporary matrix with the same size of $S_{BL}$ and $S_{BL}$, respectively. 
For better readability we underline output matrices in the HLACs (lines 3 and 5).
  %\p{underline in lines 2 and 4 missing. Intentional?} 
}
    \label{fig:cholalgo}
\end{figure}

\begin{figure}
\begin{lstlisting}[style=la, numbers=left, numbersep=-8pt, numberstyle=\tiny\color{gray}]
  for(j=i+$\nu$; j<m; j+=$\nu$) {
      $U_{BL}^T*\underline{U}_{BR}(:, j:j+\nu) = S_{BR}(:,  j:j+\nu)$;
  }
\end{lstlisting}
\caption{LA code synthesized for the HLAC in line 5 of Fig.~\ref{fig:cholalgo}.}
\label{fig:trsmalgo}
\end{figure}

The HLAC corresponding to the linear system needs to be further lowered.
Since one dimension is already of size $\nu$, the HLAC is partitioned along the other dimension
(the columns of submatrices $U_{BR}$ and $S_{BR}$). \thesys{} synthesizes the
algorithm in Fig.~\ref{fig:trsmalgo}, which is inlined in the algorithm in Fig.~\ref{fig:cholalgo}
replacing line 5. Now, the computation of the HLAC in~\eqref{eq:runex} is reduced to computations
on sBLACs and HLACs of size $\nu \times \nu$; \thesys{} proceeds by generating the corresponding vector-size codelet
for the latter.

\begin{figure}
\includegraphics[width=.95\columnwidth]{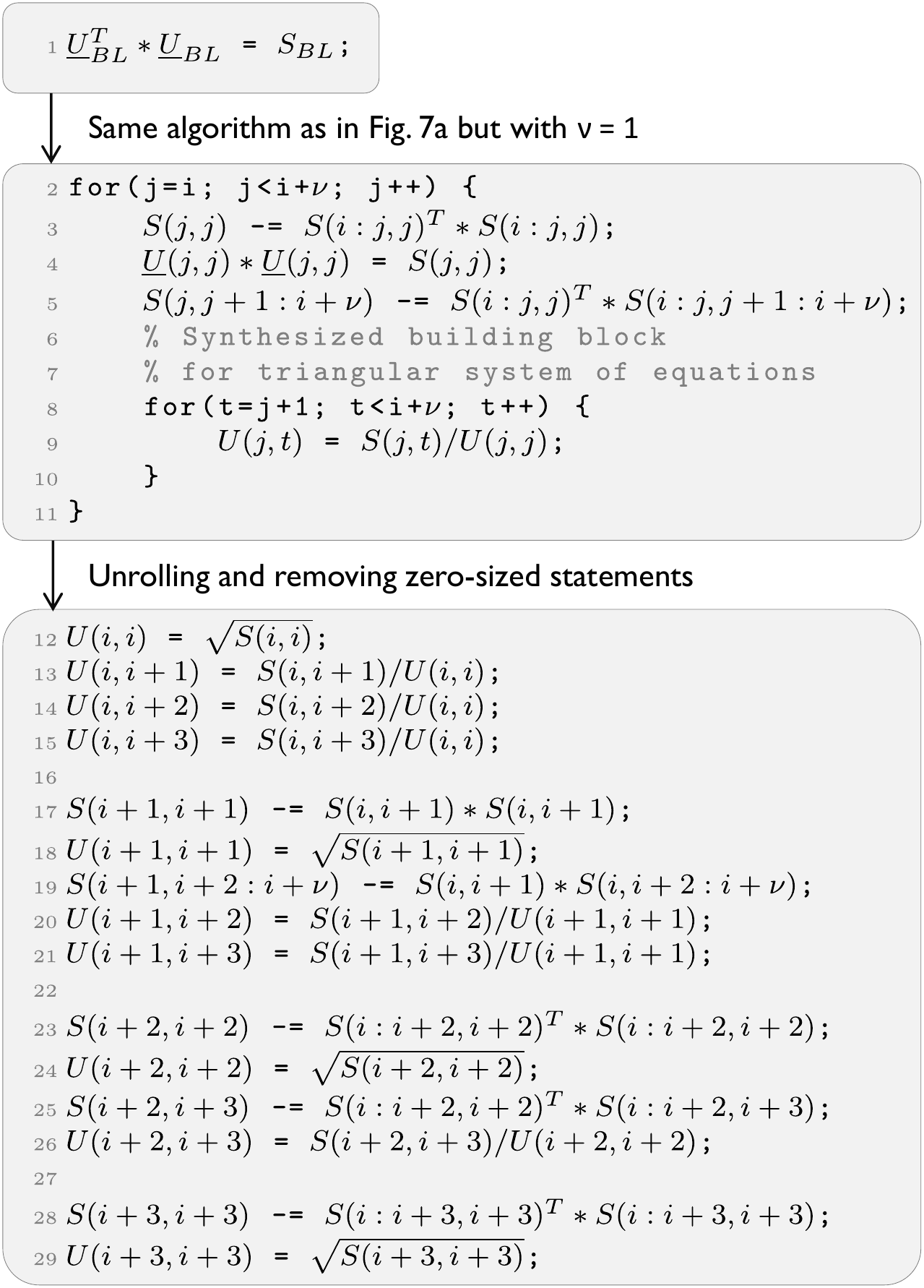} 
\caption{Code synthesis for the vector-sized HLAC in line 3 of Fig.~\ref{fig:cholalgo}. We denote with $i:j$ the interval $[i,j)$.}\label{fig:cholbldblk}.
\vspace{-1em}
\end{figure}

%\paragraph{2) Automatic synthesis of HLAC building blocks.}
\paragraph{2) Automatic synthesis of HLAC codelets.}
The remaining task is to generate code for the vector-sized HLACs; we use line 3 of
Fig.~\ref{fig:cholalgo} as example. Using the same process as before but with block size 1,
\thesys{} obtains the algorithm in Fig.~\ref{fig:cholbldblk}, lines 2--11.
This algorithm is then unrolled, as shown in lines 12--29,
and then inlined.

No further HLACs remain and \thesys{} is ready to proceed with optimizations
and code generation.

\paragraph{Algorithm reuse.}
We have discussed how the iterative process of building basic linear algebra programs out of an initial LA program
can require multiple algorithmic synthesis steps per HLAC in input.
If two HLACs only differ in the sizes of their inputs and outputs but share the same functionality (\eg{}, both
solve a system $L^TX=B$ for $X$, where $L$ is lower triangular) their LA formulations would
be based on the same algorithms. Such reuse can occur even within the same HLAC, as we have shown for the case of~\eqref{eq:runex} where the building block in Fig.~\ref{fig:cholbldblk} was created based on the same algorithm initially derived for the whole computation shown in Fig.~\ref{fig:cholalgo}.
For this reason \thesys{} stores information about the algorithms required for building basic linear algebra forms
of HLACs in a database that is queried before starting a new algorithmic synthesis step (Stage 1a in Fig.~\ref{fig:arch}).

%\subsection{Step 2: Math rewriting and code generation }
\subsection{Stage 2: sBLAC tiling and vectorization}\label{sec:vec}

The aim is to lower the basic linear algebra programs produced in the
previous stage to C-IR form. To do this, \thesys{} decomposes sBLACs into
vectorizable codelets ($\nu$-BLACs), and performs optimizations to increase
vectorization efficiency when possible.

In a basic linear algebra program, every statement is either an auxiliary
scalar computation or an sBLAC. \thesys{} proceeds first by
tiling all sBLACs and decomposing them into vector-size sBLACs that can be
directly mapped onto $\nu$-BLACs, using the \lgen{} approach described in
Sec.~\ref{sec:lgen}. 
Next it improves the vectorization efficiency of the resulting
implementation. For instance, the codelet in Fig.~\ref{fig:cholbldblk} is composed of sBLACs
that can be mapped to vectorized $\nu$-BLACs, but also of several scalar
computations that could result in a low vectorization efficiency. Thus,
\thesys{} searches for opportunities to combine multiple scalar operations
of the same type into one single sBLAC with
a technique similar to the one used to identify superword-level parallelism~\cite{Larsen:00}.

For instance consider the pair of rules $R_0$ and $R_1$ in Table~\ref{tab:r01}.
$R_0$ combines two scalar divisions into an element-wise division of a vector
by a scalar,
while $R_1$ transforms such an element-wise division 
into a scalar division followed by the scaling of a vector.  The
application of rules $R_0$ and $R_1$ to lines 13--15 and 20--21 in
Fig.~\ref{fig:cholbldblk} yields two additional sBLACs for the multiplication
of a scalar times a vector as shown in Fig.~\ref{fig:llcodeplusnublac}.  Similar rules 
for other basic operators create new sBLACs to improve code
vectorization.

The basic linear algebra programs are now mapped onto $\nu$-BLACs and
translated into C-IR code.

\begin{table}
\footnotesize
\caption{Example of rewriting rules to expose more $\nu$-BLACs. $x, b\in\R{}^{k}$; $\beta_i, \chi_i, \lambda,\tau\in\R{}$ . Statement $S_0$ appears in the computation before $S_1$ and no operation writes to $\chi_1$, $\beta_1$, or $\lambda$ in between.}
\label{tab:r01}
\begin{align}
%R_0&: \frac{S_0: X_0 = U^{-1}B_0,\quad S_1: X_1 = U^{-1}B_1}{X = [X_0\mid X_1], B = [B_0\mid B_1], X = U^{-1}B} \\
R_0&: \frac{S_0: \chi_0 = \beta_0/\lambda,\quad S_1: \chi_1 = \beta_1/\lambda}%
        {\texttt{x} = [\chi_0\mid \chi_1], \texttt{b} = [\beta_0\mid \beta_1], \texttt{x} = \texttt{b}/\lambda} \\
R_1&: \frac{\texttt{op(x) = op(b)/}\lambda,\quad \texttt{op(}\cdot\texttt{) = (}\cdot\texttt{) or }\cdot^\texttt{T}}{\tau\texttt{ = 1/}\lambda,\quad \texttt{op(x) = }\tau\texttt{ * op(b)}}
\end{align}
%\caption{Example of rewriting rules to expose more $\nu$-BLACs. $X_i, B_i\in\R{}^{m\times n_i}$; $U\in\R{}^{m\times m}$;  $X, B\in\R{}^{m\times (n_0+n_1)}$; $x, b\in\R{}^{k}$; and $\lambda,\tau\in\R{}$ . Statement $S_0$ appears in the computation before $S_1$ and no operation writes to $X_1$, $B_1$, or $U$ in between.}
%\vspace{-1.5em}
\end{table}

\newsavebox{\lstsubsa}
\begin{lrbox}{\lstsubsa}
\begin{minipage}{.5\columnwidth}
\begin{lstlisting}[style=la]
$\tau_0 = 1 / U(i,i)$;
$U(i,i+1:i+\nu) =$
	$ \tau_0 S(i,i+1:i+\nu)^T$;
\end{lstlisting}
%\vspace{2mm}
\end{minipage}
\end{lrbox}

\newsavebox{\lstsubsb}
\begin{lrbox}{\lstsubsb}
\begin{minipage}{.5\columnwidth}
\begin{lstlisting}[style=la]
$\tau_1 = 1 / U(i+1,i+1)$;
$U(i+1,i+2:i+\nu) =$
	$ \tau_1 S(i+1, i+2:i+\nu)^T$;
\end{lstlisting}
%\vspace{2mm}
\end{minipage}
\end{lrbox}

\begin{figure}
\subfloat[]{\usebox\lstsubsa \label{subfig:subs1}}
\subfloat[]{\usebox\lstsubsb \label{subfig:subs2}}
\caption{Application of rules in Table~\ref{tab:r01} to~\protect\subref{subfig:subs1} lines 13--15  and~\protect\subref{subfig:subs2} lines 20--21 in the code in Fig.~\ref{fig:cholbldblk}, which yields additional $\nu$-BLACs (second line of both~\protect\subref{subfig:subs1} and~\protect\subref{subfig:subs2}).}
\label{fig:llcodeplusnublac}
\end{figure}

\subsection{Stage 3: Code-level optimization and autotuning}\label{sec:locality}

In the final stage, \thesys{} performs optimizations on the C-IR code generated by the previous stage. These are similar to, or extended version of those done in \lgen{}~\cite{Spampinato:16}. We focus on one extension: an improved scalarization of vector accesses enabled by a domain-specific load/store analysis. The goal of
the analysis is to replace explicit memory loads and stores by shuffles in registers in the final
vectorized C code. The technique is domain-specific as memory pointers are
associated with the mathematical layout. We explain the approach with an
example. Consider the access $S(i:i+2,i+2)$ in Fig.~\ref{fig:cholbldblk}, line
23. The elements gathered from $S$ were computed in lines 13--15 and 20-21, which were rewritten to the code in Fig.~\ref{fig:llcodeplusnublac} as explained before. Note that in this computation $U$ overwrites $S$ (see specification in Fig.~\ref{fig:laprog}, line 4). 
The associated C-IR store/load sequence for $i=0$ is
shown in Fig.~\ref{fig:slseq}, and would yield the AVX code in
Fig.~\ref{fig:woscarep}.  However, by analyzing their
overlap, we can deduce that element 1 of the first vector (\cit{smul19a}) goes
into element 0 of the result one, while element 0 of the second vector
(\cit{smul19b}) into element 1. This means the stores/loads can be replaced by a blend instruction in registers as shown in Fig.~\ref{fig:wscarep}.
 
\newsavebox{\lstbeforesle}
\begin{lrbox}{\lstbeforesle}
\begin{minipage}[b]{\columnwidth}
\begin{lstlisting}[style=customcpp]
// Store sca. mul. in Fig. 9a. U overwrites S. 
Vecstore(S+1, smul9a, [0, 1, 2], hor);
...
// Store sca. mul. in Fig. 9b. U overwrites S. 
Vecstore(S+6, smul9b, [0, 1], hor);
...
// Load from Fig. 12, l.23 
__m256d vS02_vert = Vecload(S+2, [0, 1], vert);
\end{lstlisting}
\end{minipage}
\end{lrbox}

\begin{figure}
\usebox{\lstbeforesle}
\caption{\cir{} code snippet for the access $S(i:i+2,i+2)$ with $i=0$ on line 23 of Fig.~\ref{fig:cholbldblk}.
\texttt{Vecstore(addr, var, [$p_0$,$p_1$,...], hor/vert)} (and analogous Vecload) is a C-IR vector instruction with the following meaning: Store vector variable \texttt{var} placing the element at position \texttt{$p_i$} at the ith memory location starting from address \text{addr} in horizontal (vertical) direction. }
\label{fig:slseq}
\end{figure}

\begin{figure}
\footnotesize
\centering
\vspace{3pt}
\subfloat[]{\includegraphics[width=0.49\columnwidth]{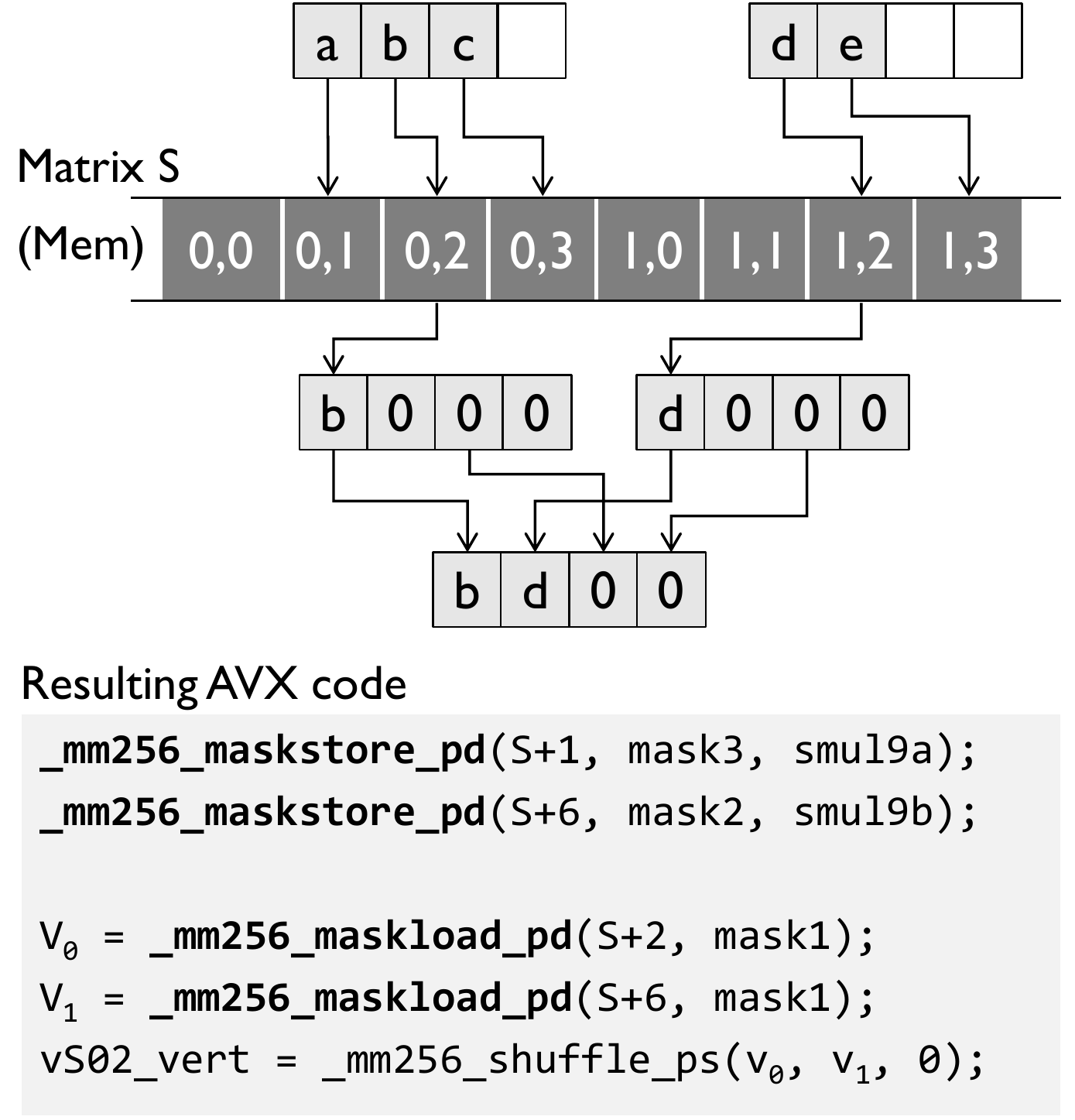}\label{fig:woscarep}}
\subfloat[]{\includegraphics[width=0.48\columnwidth]{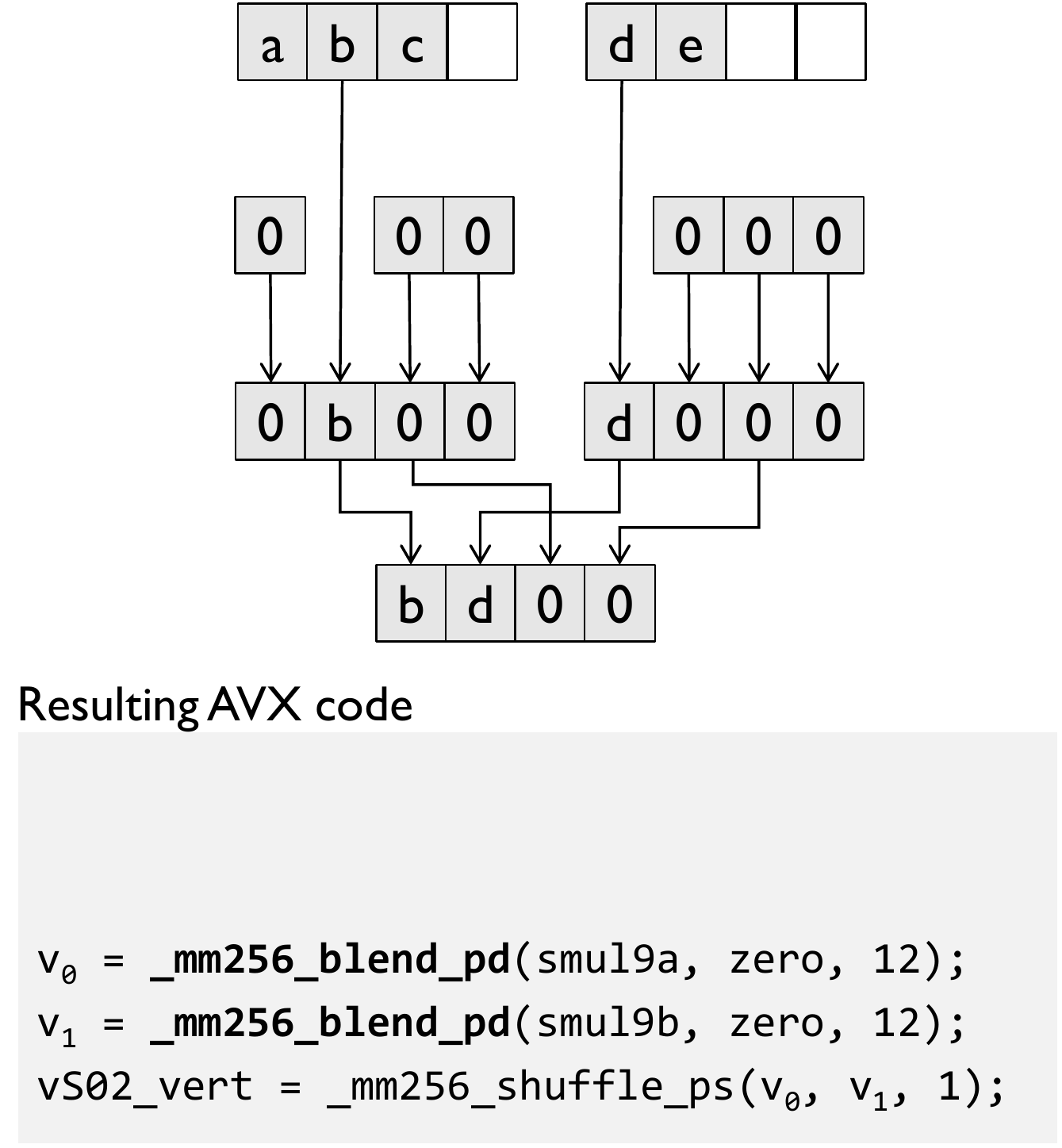}\label{fig:wscarep}}
\caption{Resulting AVX code for the \cir{} snippet in Fig.~\ref{fig:slseq} without (a) and with (b) load/store analysis. In (a)
\cit{vS02_vert} is obtained by explicitly storing to and loading from memory while in (b) by shuffling vector variables.}
\label{fig:scarep}
\end{figure}

\paragraph{Autotuning.}
Finally, \thesys{} unparses the optimized \cir{} code into C code and its performance is measured. If several algorithms are available for the occurring HLACs, autotuning is used to select the fastest.
%At this point, if more algorithms are available, \thesys{} resumes from Step 2 in Figure~\ref{fig:arch} and the fastest code is provided as an output.

\section{Experimental Results}
\label{sec:results}

\begin{table}
\centering
\setlength{\tabcolsep}{3pt}
\ra{1.3}
\small
    \caption{Selected HLAC benchmarks. All matrices $\in\R{}^{n\times n}$. 
    $A$ is symmetric positive definite, $S$ is symmetric, and $L$ and $U$ are lower and upper triangular, respectively.
    $X$ is the output in all cases.
    }
\label{tab:hlac}
\begin{tabular}{@{}lllll@{}}
\toprule
{Name }&{Label}&{Computation}\\
\midrule
    Cholesky dec. &\chol{} &$X^TX=A$, \quad $X$ upper triangular\\
\midrule
    Sylvester eq. & \sylv{} &$LX+XU=C$, \quad$X$ general matrix \\
\midrule
    Lyapunov eq. &\lyap{} &$LX+XL^T=S$, \quad$X$ symmetric \\
\midrule
    Triangular inv. &\tinv{} &$X = L^{-1}$, \quad$X$ lower triangular \\
\bottomrule
\end{tabular}
\end{table}

We evaluate \thesys{} for two classes of computations: HLACs and linear algebra applications (see Fig.~\ref{fig:compclasses}).

\paragraph{HLACs.} We selected four HLACs common in applications: the Cholesky
decomposition (\chol{}), the solution of triangular, continuous-time Sylvester and Lyapunov equations (\sylv{} and \lyap{}, respectively), and the inverse of a triangular matrix (\tinv{}). We provide their definitions in Table~\ref{tab:hlac}.

\paragraph{Applications.} We selected three applications from different domains: (a) The Kalman filter (\kf{}) for control systems which was introduced in Sec.~\ref{sec:intro}, (b) a version of the Gaussian process regression~\cite{Rasmussen:05} (\gpr{}) used in machine learning to compute the predictive mean and variance for noise free test data, and (c) an 
L1-analysis convex solver~\cite{Becker:11} (\lone{}) used, for example, in image denoising and text analysis. 
We list the associated LA programs in Fig.~\ref{fig:laprogs}.

\newsavebox{\lakf}
\begin{lrbox}{\lakf}
\begin{minipage}{.3\textwidth}
\begin{lstlisting}[style=la]
$\text{\bf Input: } F, B, Q, H, R, P, u, x, z$
$\text{\bf Output: } P, x$

$y = F*x + B*u$;
$Y = F*P*F^T + Q$;
$v_0 = z - H*y$;
$M_1 = H*Y$;    
$M_2 = Y*H^T$;
$M_3 = M1*H^T + R$;
$\underline{U}^T*\underline{U}   = M_3$;
$U^T*\underline{v_1} = v_0$;
$U*\underline{v_2} = v_1$;
$U^T*\underline{M_4} = M_1$;
$U*\underline{M_5} = M_4$;
$x = y + M_2*v_2$;
$P = Y - M_2*M_5$;
\end{lstlisting}
\end{minipage}
\end{lrbox}

\newsavebox{\lagpr}
\begin{lrbox}{\lagpr}
\begin{minipage}{.3\textwidth}
\begin{lstlisting}[style=la]
$\text{\bf Input: } K, X, x, y$
$\text{\bf Output: } \phi, \psi, \lambda$

$\underline{L}*\underline{L}^T = K$;
$L*\underline{t_0} = y$;
$L^T*\underline{t_1} = t_0$;
$k = X*x$;
$\phi = k^T*t_1$;
$L*\underline{v} = k$;
$\psi = x^T*x - v^T*v$;
$\lambda  = y^T*t_1$;
$\phantom{X}$
$\phantom{X}$
$\phantom{X}$
$\phantom{X}$
$\phantom{X}$
\end{lstlisting}
\end{minipage}
\end{lrbox}

\newsavebox{\lalone}
\begin{lrbox}{\lalone}
\begin{minipage}{.3\textwidth}
\begin{lstlisting}[style=la]
$\text{\bf Input: } W, A, x_0, y, v_1, z_1, v_2, z_2, \alpha, \beta, \tau$
$\text{\bf Output: } v_1, z_1, v_2, z_2$

$y_1 = \alpha*v_1 + \tau*z_1$;
$y_2 = \alpha*v_2 + \tau*z_2$;
$x_1 = W^T*y_1 - A^T*y_2$;
$x  = x_0 + \beta*x_1$;
$z_1 = y_1 - W*x$;
$z_2 = y_2 -(y - A*x)$;
$v_1 = \alpha*v_1 + \tau*z_1$;
$v_2 = \alpha*v_2 + \tau*z_2$;
$\phantom{X}$
$\phantom{X}$
$\phantom{X}$
$\phantom{X}$
$\phantom{X}$
\end{lstlisting}
\end{minipage}
\end{lrbox}

\begin{figure*}
\subfloat[Program \kf{}]{\usebox\lakf \label{subfig:lakf}}
\subfloat[Program \gpr{}]{\usebox\lagpr \label{subfig:lagpr}}
\subfloat[Program \lone{}]{\usebox\lalone \label{subfig:lalone}}
\caption{Selected application benchmarks. The declaration of input and output
  elements is omitted but we underline output matrices and vectors in all
  HLACs. All matrices and vectors are of size $n\times n$ and $n$,
  respectively. Both \kf{} and \lone{} are iterative algorithms and we limit
  our LA implementation to a single iteration. The original algorithms of both
  \gpr{} and \lone{} contain additionally a small number of min, max, and log operations, which have very minor impact on the overall cost. }\label{fig:laprogs}
\end{figure*}

\subsection{Experimental setup}
All tests are single-threaded and executed on an Intel Core i7-2600 CPU (Sandy Bridge) running at 3.3 GHz, with 32 kB L1 D-cache, 256 kB L2 cache, and support for AVX, under Ubuntu 14.04 with Linux kernel v3.13. Turbo Boost is disabled. In the case of \chol{}, \sylv{}, \lyap{}, and \tinv{} we compare with: (a) the Intel MKL library v11.3.2, (b) ReLAPACK~\cite{Peise:16}, (c) Eigen v3.3.4~\cite{Eigen}, straightforward code (d) compiled with Intel icc v16, and (e) clang v4 with the polyhedral Polly optimizer~\cite{Grosser:12}, and (f) the implementation of algorithms generated by \click{} implemented with MKL. For \sylv{} we also compare with RECSY~\cite{Jonsson:02}, a library specifically designed for these solvers. In the case of \kf{}, \gpr{}, and \lone{} we compare against library-based implementations using MKL and Eigen. Note that starting with v11.2, Intel MKL added specific support for small-scale, double precision matrix multiplication ({\em dgemm}). 

The straightforward code is scalar, handwritten, loop-based code with hardcoded sizes of the matrices. It is included to show optimizations performed by the compiler. 
For icc we use the flags -O3 -xHost -fargument-noalias -fno-alias -no-ipo -no-ip. 
Tests with clang/Polly were compiled with flags -O3 -mllvm -polly -mllvm -polly-vectorizer=stripmine.  
Finally, tests with MKL are linked to the sequential version of the library using flags
from the Intel MKL Link Line Advisor.\footnote{\url{software.intel.com/en-us/articles/intel-mkl-link-line-advisor}}
In Eigen we used fixed-size Map interfaces to existing arrays, no-alias assignments, in-place computations of solvers, and
enabled AVX code generation. All code is in double precision and the working set fits in the first two levels of cache.

\paragraph{Plot navigation.}
The plots present performance in flops per cycles ($f/c$) on the $y$-axis and
the problem size on the $x$-axis. All matrices and vectors are of size $n\times n$ and $n$, respectively. The peak performance of the CPU is 8 $f/c$. A bold black line represents the fastest \thesys{}-generated code. For HLACs, generated alternative based on different \click{}-generated algorithms are shown using colored, dashed lines without markers. The selection of the fastest is thus a form of algorithmic autotuning.

Every measurement was repeated 30 times on different random inputs. The median is shown and quartile information is reported with whiskers (often too small to be visible). All tests were run with warm cache.

\subsection{Results and analysis}

\begin{figure}
\centering
\subfloat[\chol{} - Cost $\approx n^3/3$ flops]{\includegraphics[width=0.5\columnwidth]{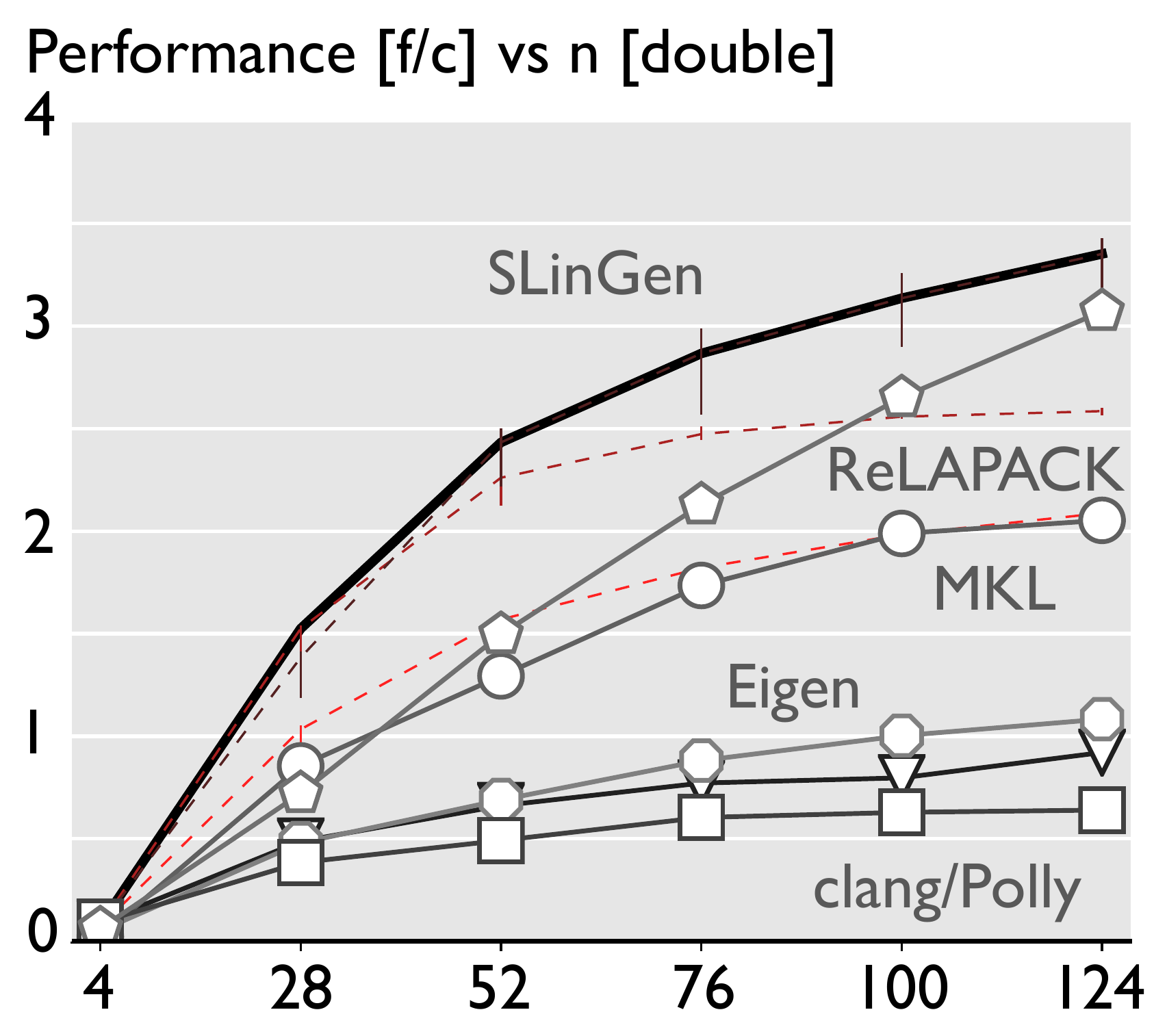}\label{fig:chol}
\includegraphics[width=0.5\columnwidth]{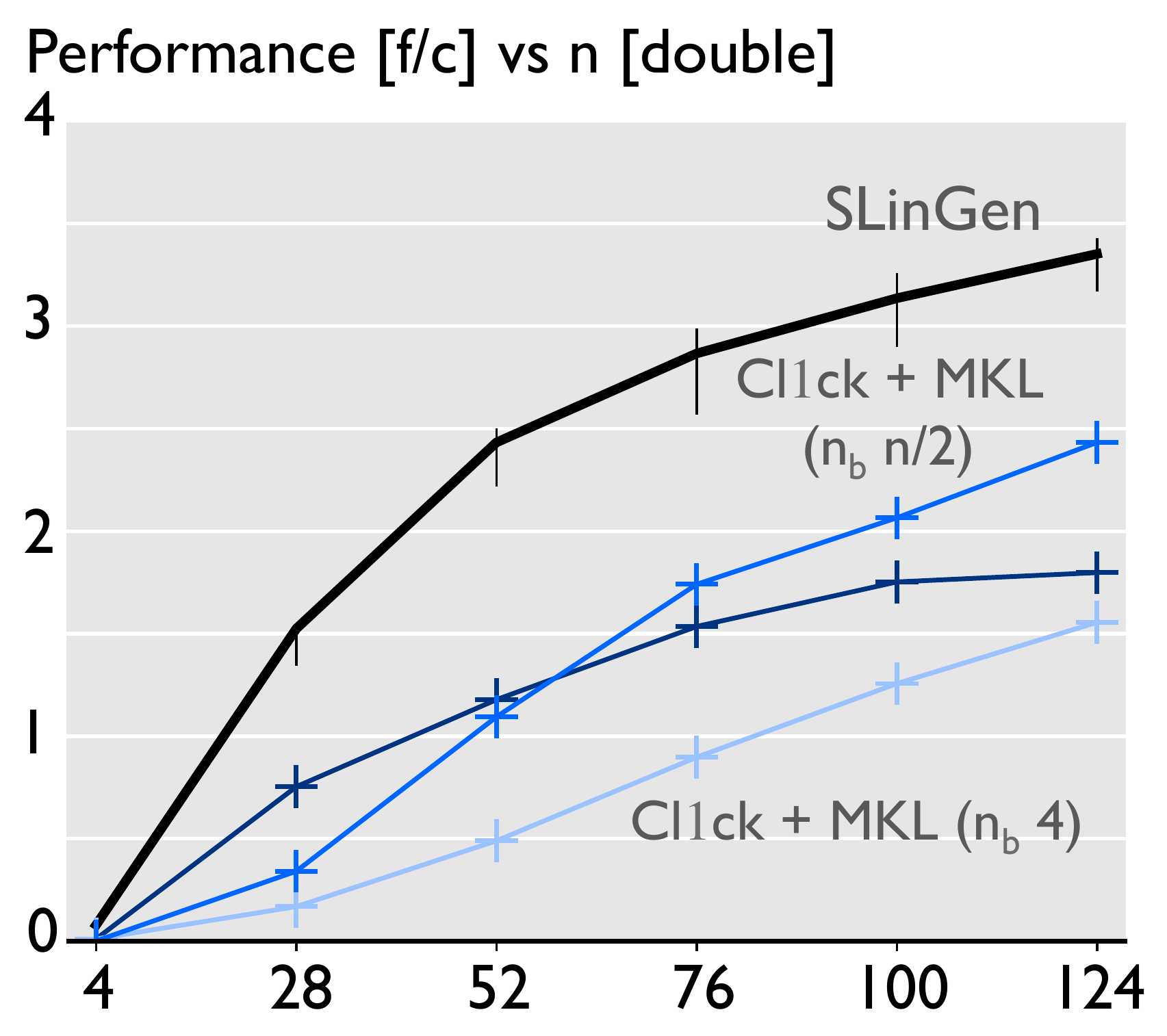}\label{fig:cholclk}}\\
\subfloat[\sylv{} - Cost $\approx 2n^3$ flops]{\includegraphics[width=0.5\columnwidth]{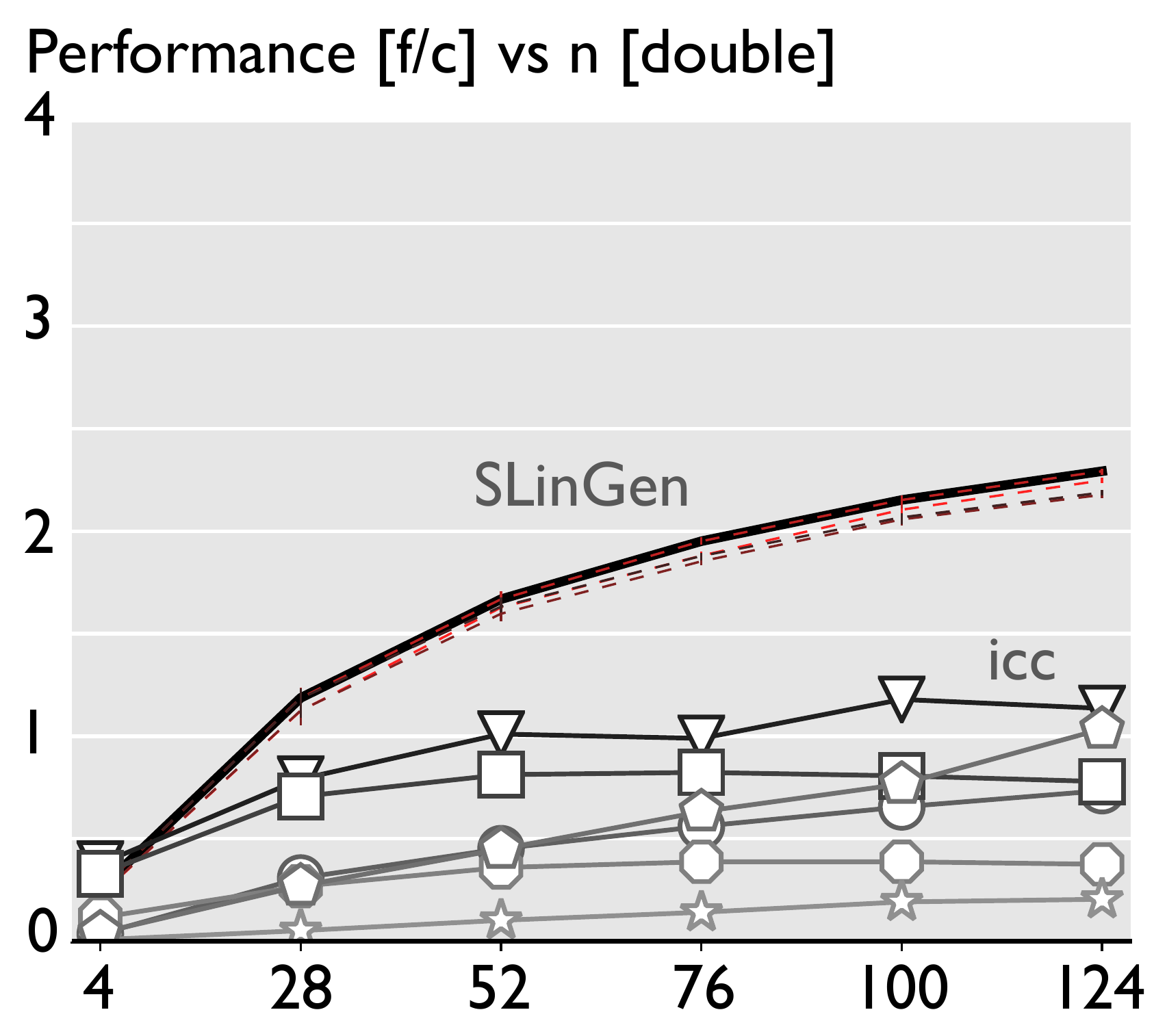}\label{fig:sylv}\includegraphics[width=0.5\columnwidth]{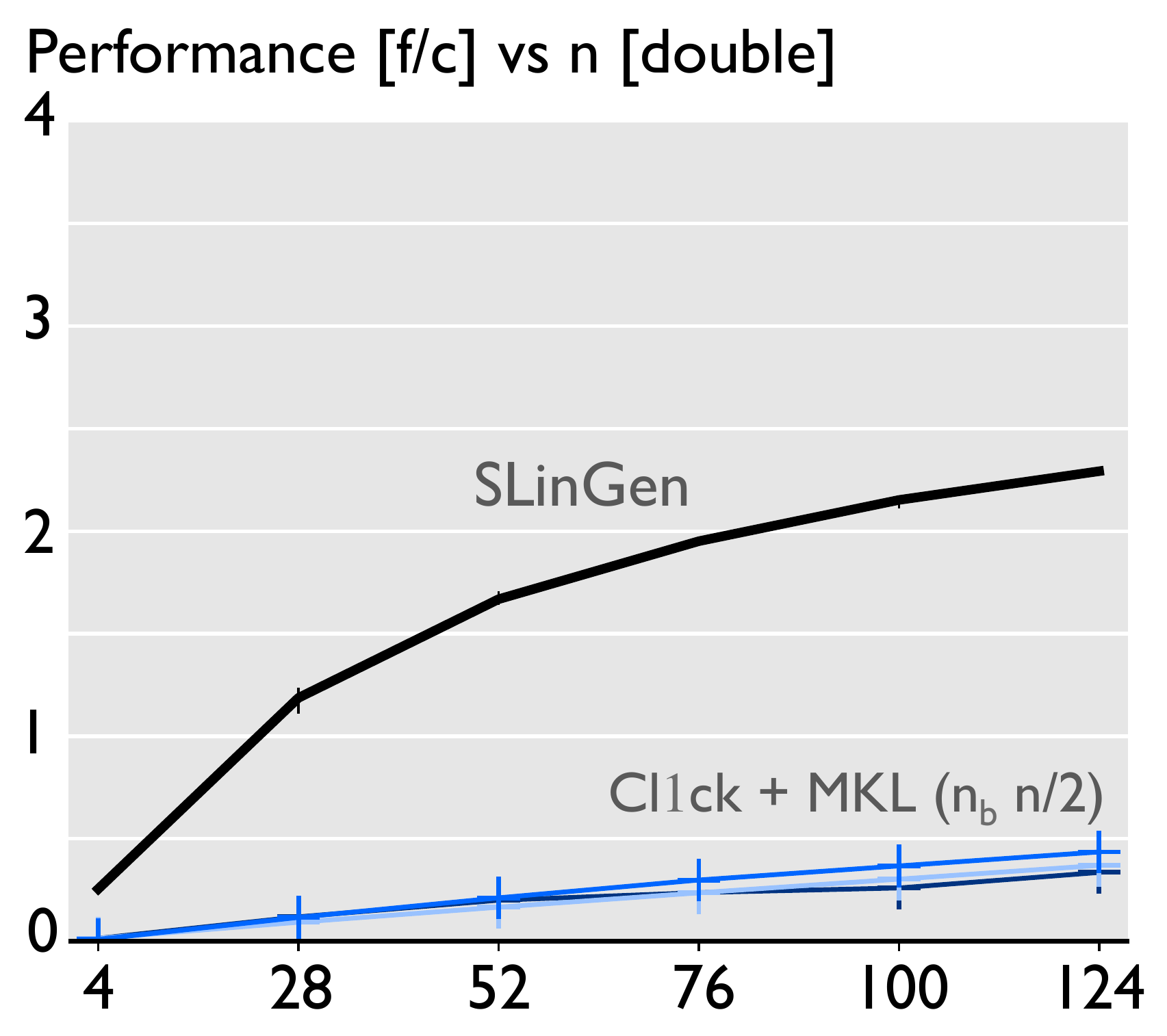}\label{fig:cholclk}}\\
\subfloat[\lyap{} - Cost $\approx n^3$ flops]{\includegraphics[width=0.5\columnwidth]{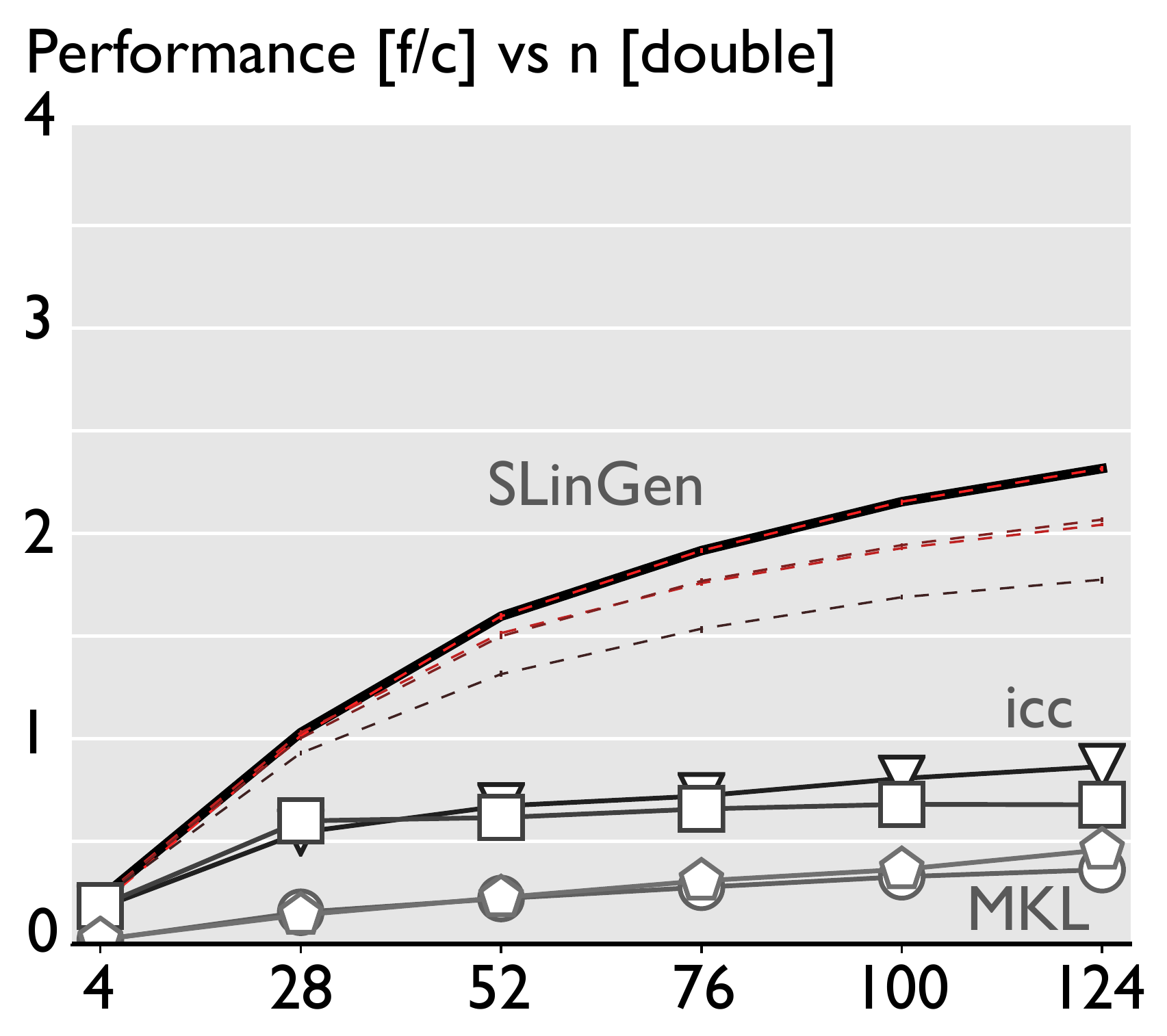}\label{fig:lyap}\includegraphics[width=0.5\columnwidth]{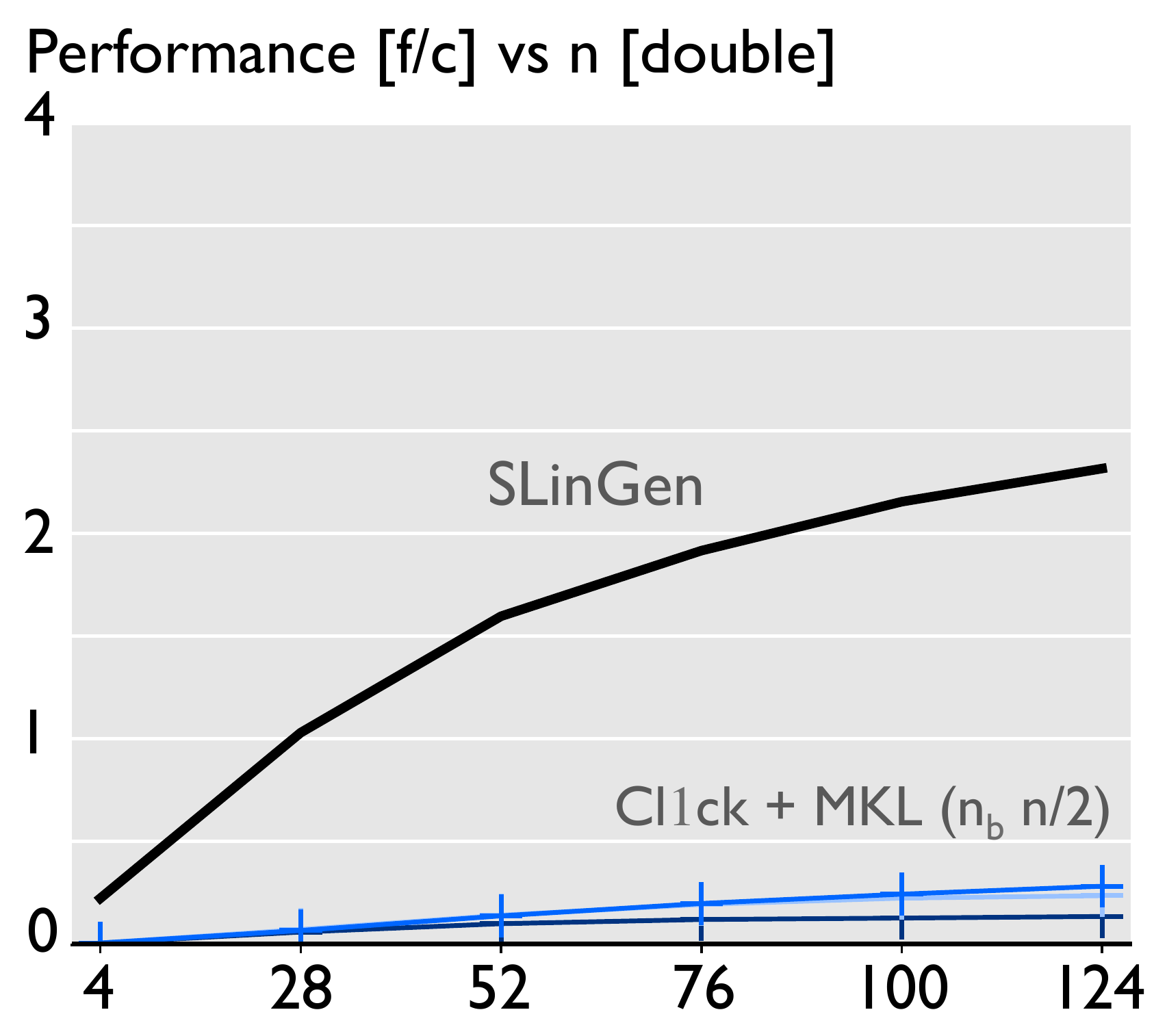}\label{fig:cholclk}}\\
\subfloat[\tinv{} - Cost $\approx n^3/3$ flops]{\includegraphics[width=0.5\columnwidth]{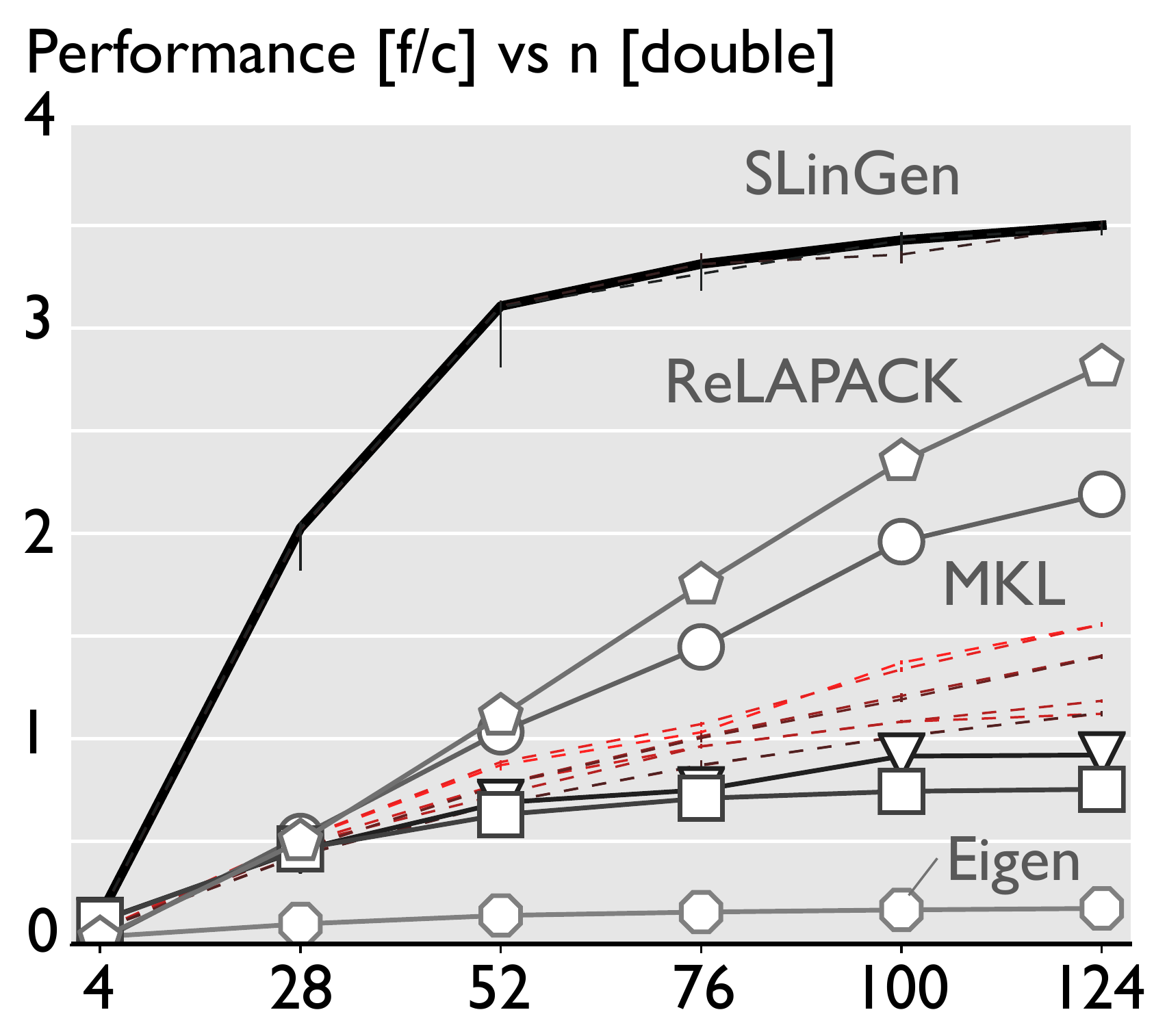}\label{fig:trinv}\includegraphics[width=0.5\columnwidth]{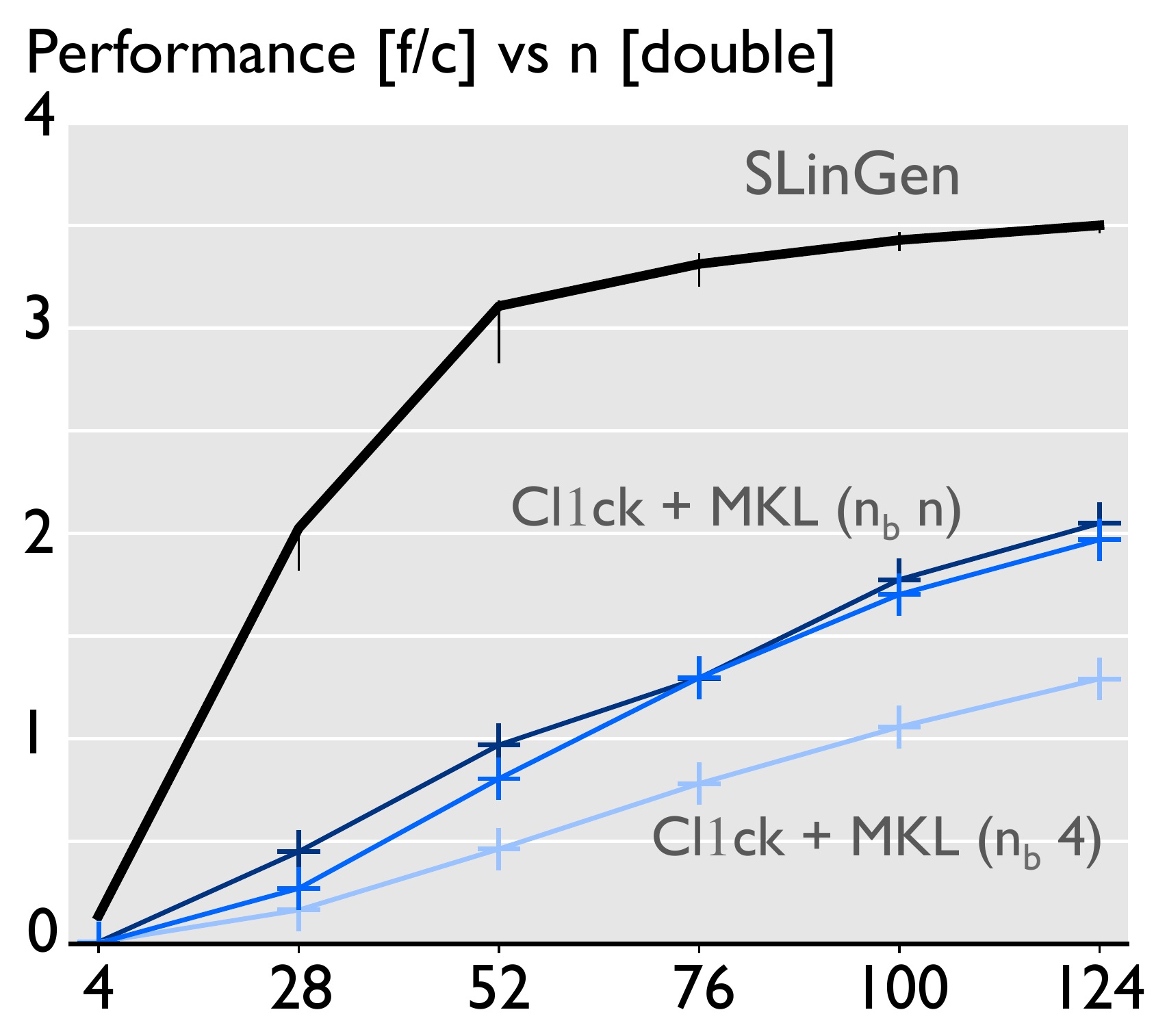}\label{fig:cholclk}}\\
\includegraphics[width=.9\columnwidth]{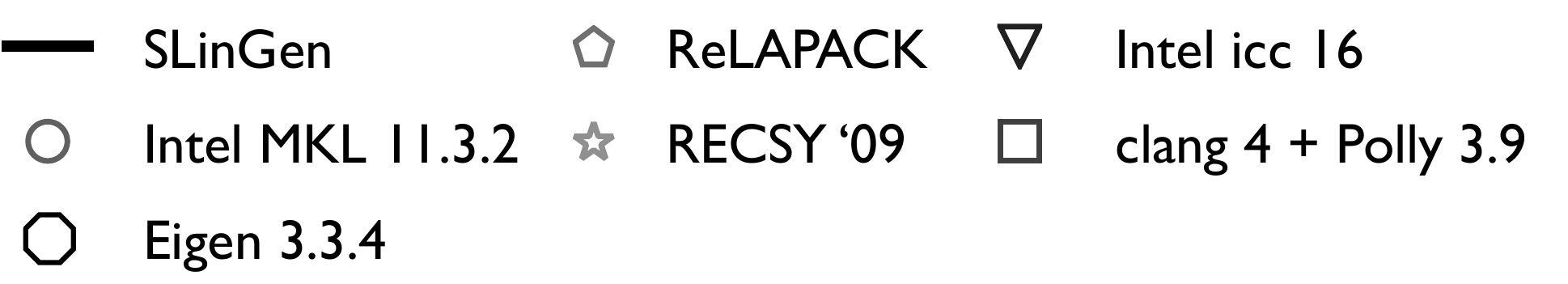}\\\vspace{1em}
\includegraphics[width=\columnwidth]{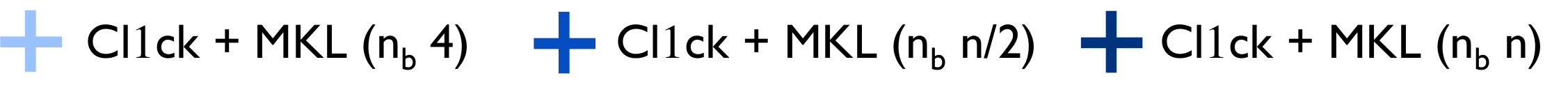}
    \caption{Performance plots for the HLAC benchmarks: \protect\subref{fig:chol} \chol{}, \protect\subref{fig:sylv} \sylv{}, \protect\subref{fig:lyap} \lyap{}, and \protect\subref{fig:trinv} \tinv{}. On the left column: colored, dashed lines without markers indicate different \thesys{}-generated algorithms.
}
\label{fig:perfplotshlac}
\end{figure}

\paragraph{HLACs.}
Figure~\ref{fig:perfplotshlac} shows the performance for the HLAC benchmarks. On the left column of Fig.~\ref{fig:perfplotshlac}, we compare performance results for all competitors except \click{}, for which we provide a more detailed comparison reported on the plots on the right column. \click{} generates blocked algorithms with a variable block size $n_b$. For each HLAC benchmark we measure performance for $n_b\in\{\nu,n/2,n\}$, where $\nu=4$ is the ISA vector length.  

For \chol{} (Fig.~\ref{fig:chol}) \thesys{} generates code that is on average $2\times$, $1.8\times$, and $3.8\times$ faster than MKL, ReLAPACK, and Eigen, respectively.  Compared to icc and clang/Polly we obtained a larger speedup of $4.2\times$ and $5.6\times$ showing the limitations of compilers. \thesys{} is also about 40\% faster than the MKL implementation of \click{}. As expected, the performance of \click{}'s implementation is very close to that of the MKL
library as this approach makes use of BLAS and LAPACK calls.
For \sylv{} (Fig.~\ref{fig:sylv}) our system reaches up
to 2 $f/c$ and is typically $2.8\times$, $2.6\times$, $12\times$, $4\times$, $1.5\times$, and $2\times$ faster than MKL, ReLAPACK, RECSY, Eigen, icc, and clang/Polly, respectively. When compared with \click{}'s implementation, \thesys\ results $5.3\times$ faster. The computation of \lyap{} (Fig.~\ref{fig:lyap}) attains around 1.7 $f/c$ for the larger sizes, thus being $5\times$ faster than the libraries and $2\times$ than the compilers. Note that missing a specialized interface for this function, MKL performs more than $2\times$ slower than icc.
Fig.~\ref{fig:trinv} shows \tinv{} where \thesys{} achieves up to 3.5 $f/c$, with an average speedup of about $2.5\times$ with respect to MKL and \click{}'s most performant implementation. When comparing with the other competitors, \thesys{} is up to $2.3\times$, $21\times$, $4.2\times$, and $4.6\times$ faster than MKL, ReLAPACK, Eigen, icc and clang/Polly, respectively.

\begin{figure}
\centering
\subfloat[\kf{} - Cost $\approx 11.3n^3$ flops]{\includegraphics[width=0.5\columnwidth]{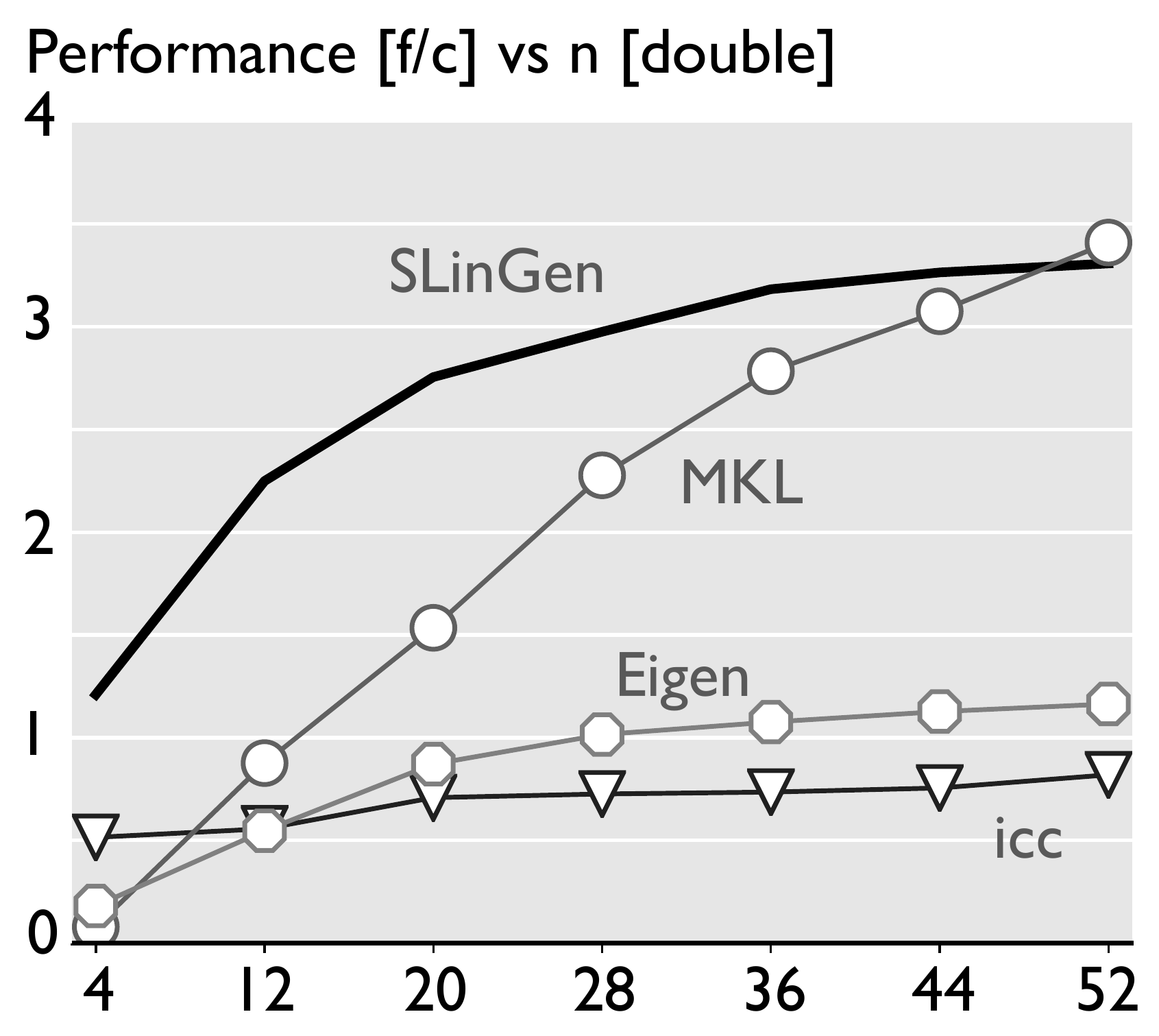}\label{fig:kf}}
\subfloat[\kf{}-28 - Cost $\approx k^3/3$ flops]{\includegraphics[width=0.5\columnwidth]{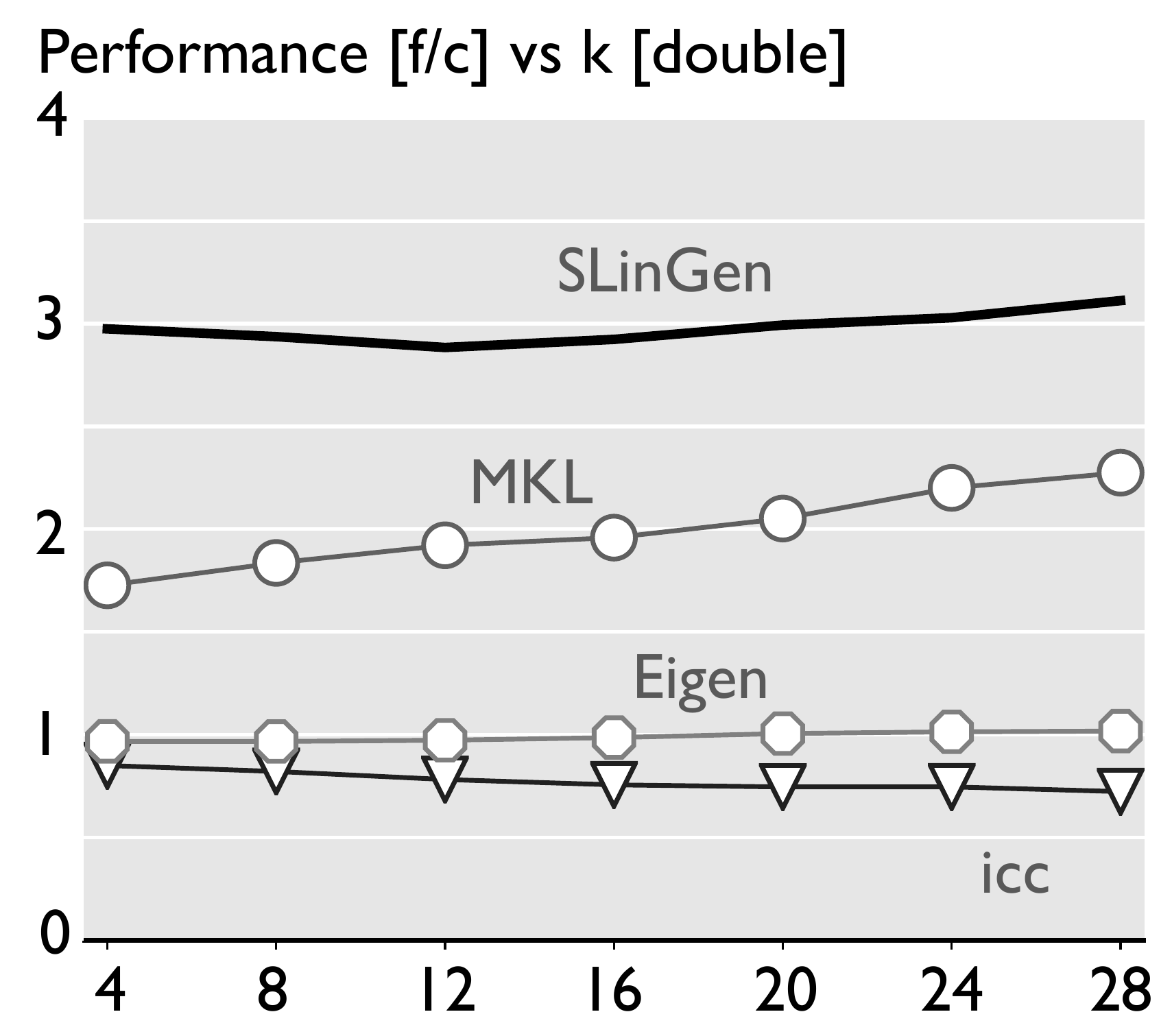}\label{fig:kf28}}\\
\subfloat[\gpr{} - Cost $\approx n^3/3$ flops]{\includegraphics[width=0.5\columnwidth]{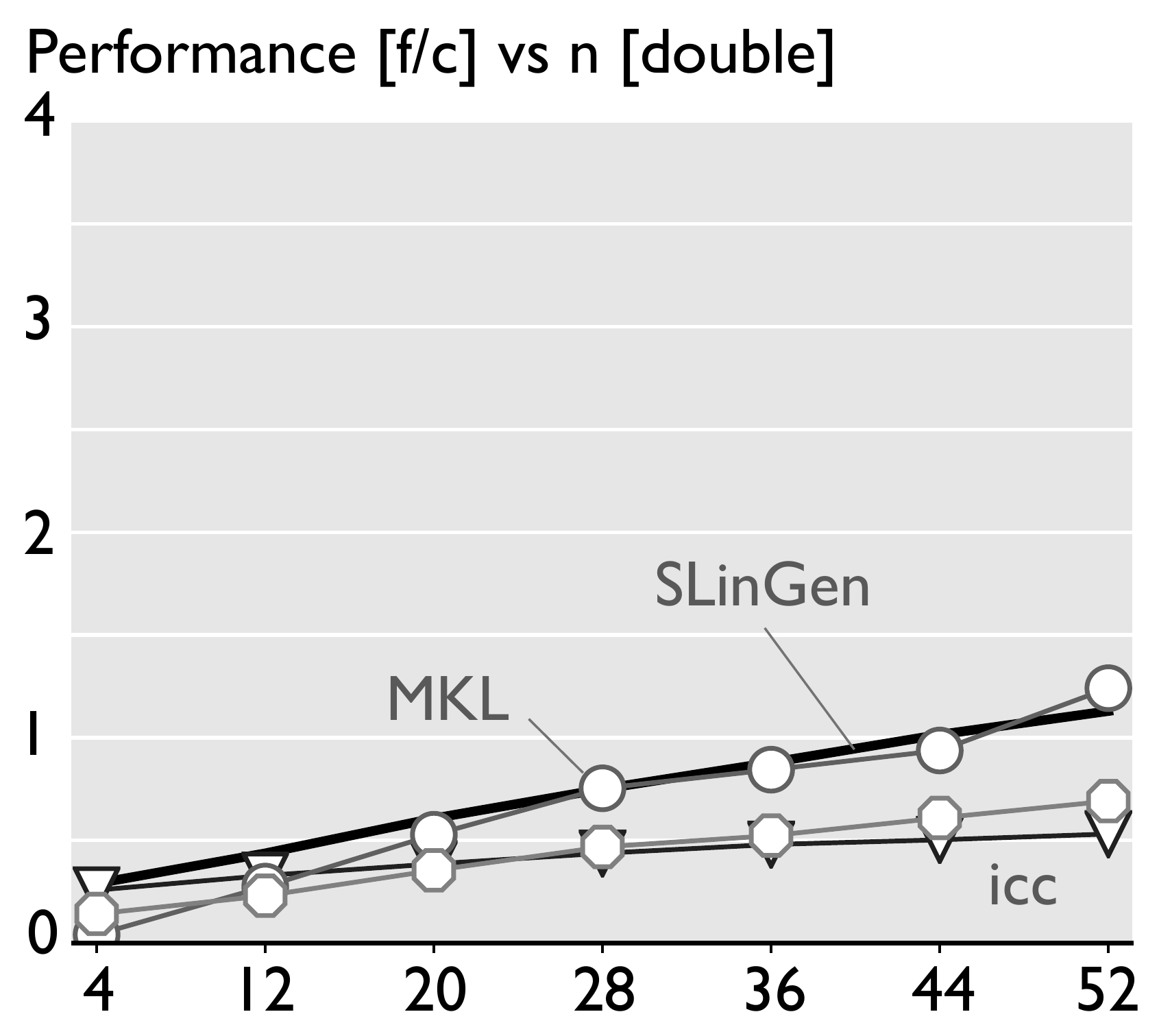}\label{fig:gpr}}
\subfloat[\lone{} - Cost $\approx 8n^2$ flops]{\includegraphics[width=0.5\columnwidth]{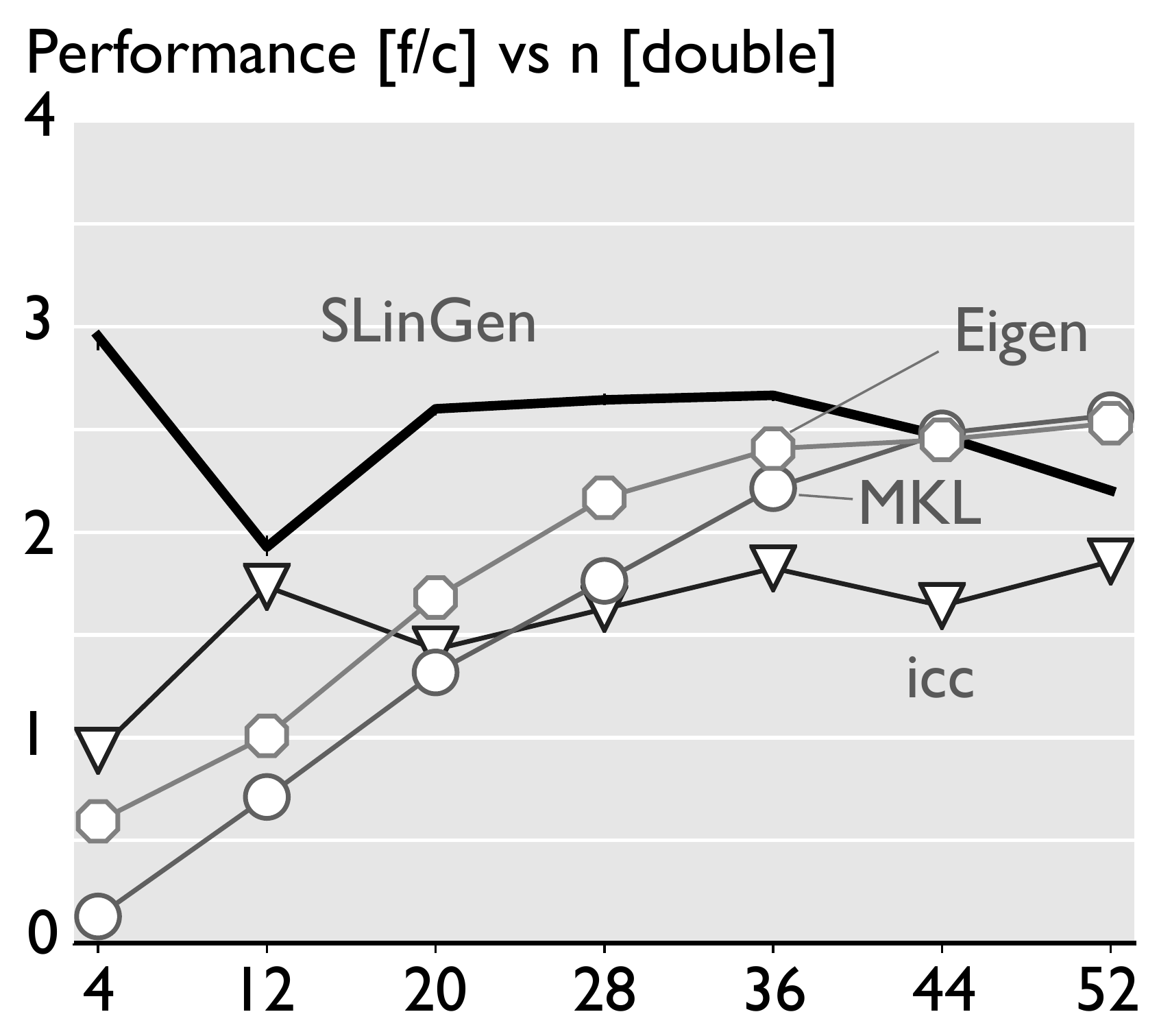}\label{fig:lone}}\\
\includegraphics[width=\columnwidth]{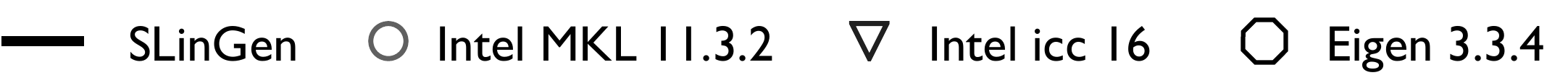}
%\begin{minipage}[t]{0.3\textwidth}
%\hspace{.06\textwidth}
%\raisebox{10pt}{\includegraphics[height=.73\textwidth]{figs/leg.pdf}}
%\end{minipage}
    \caption{Performance plots for the application-level programs in fig~\ref{fig:laprogs}. \kf{}-28 is based on the same \kf{} program but fixes the number of state parameters to 28. Inputs $z$, $H$, and $R$ have now size $k$, $k\times 28$, and 
    $k\times k$, respectively.}
\label{fig:perfplotslap}
%\vspace{-1em}
\end{figure}

\paragraph{Applications.}
Figure~\ref{fig:perfplotslap} shows the performance for the chosen applications. For \kf{}, we show two plots. Fig.~\ref{fig:kf} varies the state size (i.e., dimension of $x$ and $P$) and sets the observation size (i.e., dimension of $z$, $H$, and $R$) equal to it. \thesys{} generates code which is on average $1.4\times$, $3\times$, and $4\times$ faster than MKL, Eigen, and icc. Note that typical sizes for \kf{} tend to lie on the left half, where we observe even larger speedups. Fig.~\ref{fig:kf28} fixes the state size to 28 and varies the size of the observation size between 4 and 28.

On \gpr{}, our generated code performs similarly to MKL, while being $1.7\times$ faster than icc and Eigen. Finally, for the \lone{} test, \thesys{} generated code that is on average $1.6\times$, $1.3\times$, and $1.5\times$ faster than MKL, Eigen, and icc, respectively.

To judge the quality of the generated code, we study more in depth the code synthesized for the HLAC benchmarks.

\paragraph{Bottleneck analysis.}
We examined \thesys{}'s generated code with ERM~\cite{Caparros:14}, a tool that performs a generalized roofline analysis to determine hardware bottlenecks. ERM creates the computation DAG using the LLVM Interpreter~\cite{Lattner:04,Llvm} and a number of microarchitectural parameters capturing throughput and latency information of instructions and the memory hierarchy. We ran ERM with the parameters describing our Sandy Bridge target platform to analyze our four HLAC benchmarks. 

Table~\ref{tab:bottlenecks} summarizes for each routine and three sizes the hardware bottleneck found with ERM. In general we notice that the generated code is limited by two main factors. For small sizes it is the cost of divisions/square roots, which on Sandy Bridge can only be issued every 44 cycles. The fraction of these is asymptotically small (approximately $1/n^2$ for \chol{}, and $1/n$ for \sylv{}, \lyap{}, and \tinv{}) but matters for small sizes as they are also sequentially dependent.
We believe that this effect plagues our \gpr{}, which suffers from divisions by Cholesky decomposition and triangular solve.

For larger sizes, the number of loads and stores between registers and L1 grows mostly due to spills. \thesys{}
tends to fuse several innermost loops of several neighbouring sBLACs, which can result in increased register pressure due to several intermediate computations. This effect could be improved by better scheduling strategies based on analytical models~\cite{Low:16}.

Finally, the third column in Table~\ref{tab:bottlenecks} shows the ratio of issued shuffles and blends to the total number of issued instructions, excluding loads and stores. This number can be considered as an estimate of how much data rearrangement overhead is introduced by \thesys{}'s vectorization strategy. As estimated by ERM's bottleneck analysis, the peak performance would almost never be affected by the introduced shuffles and blends (forth and fifth column).

\begin{table}
\centering
\setlength{\tabcolsep}{3pt}
\ra{1.1}
\footnotesize
\caption{Summary of bottleneck analysis with ERM. The shuffle/blend issue rate is the ratio of shuffle/blend issues to the total issued instructions (excluding loads and stores). The achievable peak performance when taking shuffles/blends into account is shown in the last columns.}
\label{tab:bottlenecks}
\begin{tabular}{@{}llllll@{}}
\toprule
\multirow{2}{*}{{Computation}}&\multirow{2}{*}{{Sizes}}&\multirow{2}{*}{{Bottleneck}}&{Issue rate}&\multicolumn{2}{c}{Perf limit ($f/c$)}\\
&&&{Shuffles \& blends }& {Shuffles} & {Blends}\\
\midrule
\multirow{3}{*}{\chol{}}& 4 & divs/sqrt & 50\% &6.5 & 8\\
&76& L1 stores & 15\% &8 & 8 \\
&124& L1 stores & 10\% &8 & 8 \\
\midrule
\multirow{3}{*}{\sylv{}}& 4 & divs & 52\% &5.2 & 8\\
&76& divs & 35\% &8 & 8\\
&124& L1 loads & 35\% &8 & 8\\
\midrule
\multirow{3}{*}{\lyap{}}& 4 & divs & 55\% &4.5 & 8\\
&76& divs & 40\% &6.7 & 8\\
&124& L1 loads & 37\% &6.8 & 8 \\
\midrule
\multirow{3}{*}{\tinv{}}& 4 & divs & 62\% &3.2 & 8\\
&76& L1 loads & 32\% &8 & 8\\
&124& L1 loads & 32\% &8 & 8\\
\bottomrule
\end{tabular}
\end{table}

\section{Limitations and Future Work}

\thesys{} is a research prototype and naturally has limitations that future work can address. 

First, \thesys{} is currently restricted to fixed input and output sizes and expects structured matrices to 
use a full storage scheme. Many relevant applications, e.g., in signal processing, control, and optimization fulfil this constraint. However, for others a library for general input sizes is still desirable.

Second, when a linear algebra computation is composed of many small subcomputations, \thesys{} could take advantage of modern multicore systems. The idea is to identify task parallelism among independent subcomputations and allocate them to different cores of a processor.

Third, we only consider computations on a single input. However, \thesys{} could be extended to handle batched computations, \ie{}, a group of independent computations that can fit into cache together and could be processed by one single function to better take advantage of data and task parallelism. 

%Fourth, the application-level benchmarks are only the first step towards entire applications. Investigating the integration of \thesys{} into well-established mathematical environments, such as MATLAB, or computer algebra systems, such as Mathematica, could help expand to a much larger set of applications and provide ground for potential cross-domain performance optimizations.

Finally, the generated code is certainly not performance-optimal. More could be achieved by additional lower level optimizations that address the issues identified in our bottleneck analysis.

\section{Related Work}

%There is a rich body of work on optimizing linear algebra computations by hand or automatically that we review here. 
Here we review the existing body of work on the optimization of linear algebra computations, performed either by hand or automatically.

\sloppypar
\paragraph{Hand-optimized linear algebra libraries.} Multiple libraries offer high-performance implementations of BLAS and
LAPACK interfaces. Prominent examples include the commercial Intel MKL~\cite{mkl}
and the open-source OpenBLAS~\cite{OpenBLAS}.
BLIS~\cite{VanZee:15} is a framework for instantiating a superset of the BLAS operations from a set of microkernels; however, it does not does cover higher-level operations or entire applications.
ReLAPACK~\cite{Peise:16} and RECSY~\cite{Jonsson:02} provide recursive high-performance 
implementations of LAPACK and Sylvester-type computations, respectively.

\paragraph{Algorithmic Synthesis.} 
The Formal Linear Algebra Methods Environment (FLAME)~\cite{Gunnels:01b} provides a methodology for automatically deriving algorithms for higher level linear algebra functions~\cite{Bientinesi:05} given as mathematical equations. The supported functions are mostly those covered by the LAPACK library and the generated algorithms rely on the availability of a BLAS library. The methodology is completely automated by the \click{} compiler~\cite{cl1ck-pme,cl1ck-linv} which we used and extended in this work.

\paragraph{Library focus on small computations.}
Recently, the problem of efficiently processing small to medium size 
computations has attracted the attention of hardware and library developers.
The LIBXSMM library~\cite{Heinecke:16} provides an assembly code generator for small dense and sparse matrix
multiplication, specifically for Intel platforms. Intel MKL has introduced
fast general matrix multiplication (GEMM) kernels on small matrices, as well as a specific 
functionality for batches of matrices with same parameters, such as size and leading 
dimensions (GEMM_BATCH). The very recent \cite{Frison:17}
provides carefully optimized assembly for small-scale BLAS-like operations.
The techniques used in writing these kernels could be incorporated into \thesys{}.

\paragraph{DSL-based approaches.}
PHiPAC~\cite{Bilmes:97} and ATLAS~\cite{Whaley:98} are two of the earliest generators, both aiming at high-performance GEMM by parameter tuning. 
Higher-level generators for linear algebra include the CLAK
compiler~\cite{clak:2}, the DxTer system~\cite{Marker:12}, and BTO~\cite{Siek:08}. 
CLAK finds efficient mappings of matrix equations onto building blocks
from high-performance libraries such as BLAS and LAPACK.
\thesys{} could benefit from a CLAK-like high-level front-end to perform
mathematical reasoning on more involved matrix computations that require manipulation. 
DxTer transforms blocked algorithms such as those generated by \click{},
and applies transformations and refinements to output high-performance
distributed-memory implementations. BTO focuses on memory bound computations (BLAS 1-2
operations) and relies on a compiler for vectorization. LINVIEW~\cite{Nikolic:14} is a framework for incremental maintenance of analytical
queries expressed in terms of linear algebra programs. The goal of the system is to propagate within 
a (large) computation only the changes caused by (small) variations in the input matrices.

The MATLAB Coder~\cite{MatlabCoder} supports the generation of C and C++ functions from most of the 
MATLAB language but produces only scalar code without explicit vectorization.
Julia~\cite{Bezanson:17} is is a high-level dynamic language that targets scientific
computing. Linear algebra operations in Julia are mapped to BLAS and LAPACK calls.

Also related are expression template-based DSLs like the
Eigen~\cite{Eigen} and the uBLAS~\cite{uBlas} libraries. In particular, Eigen
provides vectorized code generation, supporting a variety of
functionalities including Cholesky and LU factorizations. However, libraries
based on C++ metaprogramming cannot take advantage of algorithmic or
implementation variants. Another approach based on metaprogramming is taken by the Hierarchically Tiled Arrays (HTAs)~\cite{Guo:08}, which offer data types with the ability to dynamically partition matrices and vectors, automatically handling situations of overlapping areas. HTAs priority, however, is to improve programmability reducing the amount of code required to handle tiling and data distribution in parallel programs, leaving any optimization to the programmer (or program generator).

Finally, our approach is in concept related to Spiral, a generator for the different domain of linear transforms \cite{Pueschel:05,Pueschel:11} and also uses ideas from the code generation approach of the \lgen{} compiler~\cite{Spampinato:14, Kyrtatas:15,Spampinato:16}. Spiral is a generator for linear transforms (like FFT, a very different domain) and not for the computations considered in this paper. There was an effort to extend it to linear algebra~\cite{Franchetti:09b} but was shown only for matrix-matrix multiplication. The connection between \thesys{} and Spiral is in concept: The translation from math to code and the use of DSLs to apply optimizations at a high level of abstraction.

\paragraph{Optimizing compilers.}
In Sec.~\ref{sec:results} we compare against Polly~\cite{Grosser:12}, an optimizer for the clang compiler based on the polyhedral model~\cite{Feautrier:11}. This and other techniques reschedule 
computation and data accesses to enhance locality and expose parallelization and vectorization opportunities~\cite{Bondhugula:08,Kong:13}.
Multi-platform vectorization techniques such as those in~\cite{Nuzman:06,Nuzman:11} use abstract SIMD representations making optimizations such as alignment detection portable across different architectures. 
The work in~\cite{Rapaport:15} leverages whole-function vectorization at the C level, differently from other frameworks that deviate from the core C language, such as Intel ispc~\cite{Pharr:12}. 
The scope of these and other optimizing compilers is more general than that of our generator. On the other hand, we take advantage of the specific domain to synthesize vectorized C code of higher performance.

\section{Conclusions}

This paper pursues the vision that performance-optimized code for well-defined mathematical computations should be generated directly from a mathematical representation. This way a) complete automation is achieved; b) domain knowledge is available to perform optimizations that are difficult or impossible for compilers at a high level of abstraction; c) porting to new processor architectures may be simplified. In this paper we proposed a prototype of such a system, \thesys{}, focused on small-scale linear algebra as needed in control, signal processing, and other domains. While limited in scope, it goes considerably beyond prior work on automation, that was mainly focused on library functions for BLAS or LAPACK; \thesys{} compiles entire linear algebra programs and still obtains competitive or superior performance to handwritten code. The methods we used are a combination of DSLs, program synthesis and generation, symbolic computation, and compiler techniques. There is active research on these topics, which should also provide ever better tools and language support to facilitate the development of more, and more powerful generators like \thesys{}.

\begin{acks}                            %% acks environment is optional
%                                        %% contents suppressed with 'anonymous'
%  %% Commands \grantsponsor{<sponsorID>}{<name>}{<url>} and
%  %% \grantnum[<url>]{<sponsorID>}{<number>} should be used to
%  %% acknowledge financial support and will be used by metadata
%  %% extraction tools.
%  This material is based upon work supported by the
%  \grantsponsor{GS100000001}{National Science
%    Foundation}{http://dx.doi.org/10.13039/100000001} under Grant
%  No.~\grantnum{GS100000001}{nnnnnnn} and Grant
%  No.~\grantnum{GS100000001}{mmmmmmm}.  Any opinions, findings, and
%  conclusions or recommendations expressed in this material are those
%  of the author and do not necessarily reflect the views of the
%  National Science Foundation.

Financial support from the Deutsche Forschungsgemeinschaft (DFG, German 
Research Foundation) through grants GSC~111 and BI 1533/2--1 is 
gratefully acknowledged.
\end{acks}

% Bibliography
%\bibliographystyle{natbib}
\bibliography{main}


\newcommand{\SortNoop}[1]{}
\begin{thebibliography}{46}

%%% the Web address.

\ifx \showCODEN    \undefined \def \showCODEN     #1{\unskip}     \fi
\ifx \showDOI      \undefined \def \showDOI       #1{#1}\fi
\ifx \showISBNx    \undefined \def \showISBNx     #1{\unskip}     \fi
\ifx \showISBNxiii \undefined \def \showISBNxiii  #1{\unskip}     \fi
\ifx \showISSN     \undefined \def \showISSN      #1{\unskip}     \fi
\ifx \showLCCN     \undefined \def \showLCCN      #1{\unskip}     \fi
\ifx \shownote     \undefined \def \shownote      #1{#1}          \fi
\ifx \showarticletitle \undefined \def \showarticletitle #1{#1}   \fi
\ifx \showURL      \undefined \def \showURL       {\relax}        \fi
% invisible to TeX
\providecommand\bibfield[2]{#2}
\providecommand\bibinfo[2]{#2}
\providecommand\natexlab[1]{#1}
\providecommand\showeprint[2][]{arXiv:#2}

\bibitem[\protect\citeauthoryear{Anderson, Bai, Bischof, Blackford, Demmel,
  Dongarra, Du~Croz, Greenbaum, Hammarling, McKenney, and Sorensen}{Anderson
  et~al\mbox{.}}{1999}]%
        {Anderson:99}
\bibfield{author}{\bibinfo{person}{E. Anderson}, \bibinfo{person}{Z. Bai},
  \bibinfo{person}{C. Bischof}, \bibinfo{person}{S. Blackford},
  \bibinfo{person}{J. Demmel}, \bibinfo{person}{J. Dongarra},
  \bibinfo{person}{J. Du~Croz}, \bibinfo{person}{A. Greenbaum},
  \bibinfo{person}{S. Hammarling}, \bibinfo{person}{A. McKenney}, {and}
  \bibinfo{person}{D. Sorensen}.} \bibinfo{year}{1999}\natexlab{}.
\newblock \bibinfo{booktitle}{{\em {LAPACK} Users' Guide\/}
  (\bibinfo{edition}{third} ed.)}.
\newblock \bibinfo{publisher}{Society for Industrial and Applied Mathematics}.
\newblock


\bibitem[\protect\citeauthoryear{Becker, Cand{\`e}s, and Grant}{Becker
  et~al\mbox{.}}{2011}]%
        {Becker:11}
\bibfield{author}{\bibinfo{person}{S.~R. Becker}, \bibinfo{person}{E.~J.
  Cand{\`e}s}, {and} \bibinfo{person}{M.~C. Grant}.}
  \bibinfo{year}{2011}\natexlab{}.
\newblock \showarticletitle{Templates for convex cone problems with
  applications to sparse signal recovery}.
\newblock \bibinfo{journal}{{\em Mathematical Programming Computation\/}}
  \bibinfo{volume}{3}, \bibinfo{number}{3} (\bibinfo{year}{2011}),
  \bibinfo{pages}{165}.
\newblock


\bibitem[\protect\citeauthoryear{Bezanson, Edelman, Karpinski, and
  Shah}{Bezanson et~al\mbox{.}}{2017}]%
        {Bezanson:17}
\bibfield{author}{\bibinfo{person}{J. Bezanson}, \bibinfo{person}{A. Edelman},
  \bibinfo{person}{S. Karpinski}, {and} \bibinfo{person}{V.~B. Shah}.}
  \bibinfo{year}{2017}\natexlab{}.
\newblock \showarticletitle{Julia: A Fresh Approach to Numerical Computing}.
\newblock \bibinfo{journal}{{\it SIAM Rev.}} \bibinfo{volume}{59},
  \bibinfo{number}{1} (\bibinfo{year}{2017}), \bibinfo{pages}{65--98}.
\newblock


\bibitem[\protect\citeauthoryear{Bientinesi, Gunnels, Myers,
  Quintana-Ort\'{\i}, and Geijn}{Bientinesi et~al\mbox{.}}{2005}]%
        {Bientinesi:05}
\bibfield{author}{\bibinfo{person}{P. Bientinesi}, \bibinfo{person}{J.~A.
  Gunnels}, \bibinfo{person}{M.~E. Myers}, \bibinfo{person}{E.~S.
  Quintana-Ort\'{\i}}, {and} \bibinfo{person}{R.~A. van~de Geijn}.}
  \bibinfo{year}{2005}\natexlab{}.
\newblock \showarticletitle{The science of deriving dense linear algebra
  algorithms}.
\newblock \bibinfo{journal}{{\em ACM Transactions on Mathematical Software
  (TOMS)\/}} \bibinfo{volume}{31}, \bibinfo{number}{1} (\bibinfo{year}{2005}),
  \bibinfo{pages}{1--26}.
\newblock


\bibitem[\protect\citeauthoryear{Bilmes, Asanovic, Chin, and Demmel}{Bilmes
  et~al\mbox{.}}{1997}]%
        {Bilmes:97}
\bibfield{author}{\bibinfo{person}{J. Bilmes}, \bibinfo{person}{K. Asanovic},
  \bibinfo{person}{C.~W. Chin}, {and} \bibinfo{person}{J. Demmel}.}
  \bibinfo{year}{1997}\natexlab{}.
\newblock \showarticletitle{Optimizing matrix multiply using {PHiPAC}: a
  portable, high-performance, {ANSI C} coding methodology}. In
  \bibinfo{booktitle}{{\em International Conference on Supercomputing (ICS)}}.
  \bibinfo{pages}{340--347}.
\newblock


\bibitem[\protect\citeauthoryear{Bondhugula, Hartono, Ramanujam, and
  Sadayappan}{Bondhugula et~al\mbox{.}}{2008}]%
        {Bondhugula:08}
\bibfield{author}{\bibinfo{person}{U. Bondhugula}, \bibinfo{person}{A.
  Hartono}, \bibinfo{person}{J. Ramanujam}, {and} \bibinfo{person}{P.
  Sadayappan}.} \bibinfo{year}{2008}\natexlab{}.
\newblock \showarticletitle{A Practical Automatic Polyhedral Parallelizer and
  Locality Optimizer}. In \bibinfo{booktitle}{{\em Programming Language Design
  and Implementation (PLDI)}}. \bibinfo{pages}{101--113}.
\newblock


\bibitem[\protect\citeauthoryear{Caparr{\'o}s~Cabezas and
  P{\"u}schel}{Caparr{\'o}s~Cabezas and P{\"u}schel}{2014}]%
        {Caparros:14}
\bibfield{author}{\bibinfo{person}{V. Caparr{\'o}s~Cabezas} {and}
  \bibinfo{person}{M. P{\"u}schel}.} \bibinfo{year}{2014}\natexlab{}.
\newblock \showarticletitle{Extending the Roofline Model: Bottleneck Analysis
  with Microarchitectural Constraints}. In \bibinfo{booktitle}{{\em IEEE
  International Symposium on Workload Characterization (IISWC)}}.
  \bibinfo{pages}{222--231}.
\newblock


\bibitem[\protect\citeauthoryear{Dongarra, Du~Croz, Hammarling, and
  Duff}{Dongarra et~al\mbox{.}}{1990}]%
        {Dongarra:90}
\bibfield{author}{\bibinfo{person}{J.~J. Dongarra}, \bibinfo{person}{J.
  Du~Croz}, \bibinfo{person}{S. Hammarling}, {and} \bibinfo{person}{I.~S.
  Duff}.} \bibinfo{year}{1990}\natexlab{}.
\newblock \showarticletitle{A set of level 3 basic linear algebra subprograms}.
\newblock \bibinfo{journal}{{\em ACM Transactions on Mathematical Software
  (TOMS)\/}} \bibinfo{volume}{16}, \bibinfo{number}{1} (\bibinfo{year}{1990}),
  \bibinfo{pages}{1--17}.
\newblock


\bibitem[\protect\citeauthoryear{Fabregat-Traver and
  Bientinesi}{Fabregat-Traver and Bientinesi}{2011a}]%
        {cl1ck-linv}
\bibfield{author}{\bibinfo{person}{D. Fabregat-Traver} {and}
  \bibinfo{person}{P. Bientinesi}.} \bibinfo{year}{2011}\natexlab{a}.
\newblock \showarticletitle{Automatic Generation of Loop-Invariants for Matrix
  Operations}. In \bibinfo{booktitle}{{\em International Conference on
  Computational Science and its Applications}}. \bibinfo{pages}{82--92}.
\newblock


\bibitem[\protect\citeauthoryear{Fabregat-Traver and
  Bientinesi}{Fabregat-Traver and Bientinesi}{2011b}]%
        {cl1ck-pme}
\bibfield{author}{\bibinfo{person}{D. Fabregat-Traver} {and}
  \bibinfo{person}{P. Bientinesi}.} \bibinfo{year}{2011}\natexlab{b}.
\newblock \showarticletitle{Knowledge-Based Automatic Generation of Partitioned
  Matrix Expressions}. In \bibinfo{booktitle}{{\em Computer Algebra in
  Scientific Computing}}. \bibinfo{pages}{144--157}.
\newblock


\bibitem[\protect\citeauthoryear{Fabregat-Traver and
  Bientinesi}{Fabregat-Traver and Bientinesi}{2013}]%
        {clak:2}
\bibfield{author}{\bibinfo{person}{D. Fabregat-Traver} {and}
  \bibinfo{person}{P. Bientinesi}.} \bibinfo{year}{2013}\natexlab{}.
\newblock \showarticletitle{Application-tailored Linear Algebra Algorithms: A
  Search-Based Approach}.
\newblock \bibinfo{journal}{{\em International Journal of High Performance
  Computing Applications (IJHPCA)\/}} \bibinfo{volume}{27}, \bibinfo{number}{4}
  (\bibinfo{year}{2013}), \bibinfo{pages}{425 -- 438}.
\newblock


\bibitem[\protect\citeauthoryear{Feautrier and Lengauer}{Feautrier and
  Lengauer}{2011}]%
        {Feautrier:11}
\bibfield{author}{\bibinfo{person}{P. Feautrier} {and} \bibinfo{person}{C.
  Lengauer}.} \bibinfo{year}{2011}\natexlab{}.
\newblock \bibinfo{booktitle}{{\em Encyclopedia of Parallel Computing}}.
\newblock \bibinfo{publisher}{Springer}, Chapter Polyhedron Model.
\newblock


\bibitem[\protect\citeauthoryear{Franchetti, Mesmay, Mcfarlin, and
  P\"{u}schel}{Franchetti et~al\mbox{.}}{2009}]%
        {Franchetti:09b}
\bibfield{author}{\bibinfo{person}{F. Franchetti}, \bibinfo{person}{F. Mesmay},
  \bibinfo{person}{D. Mcfarlin}, {and} \bibinfo{person}{M. P\"{u}schel}.}
  \bibinfo{year}{2009}\natexlab{}.
\newblock \showarticletitle{Operator Language: A Program Generation Framework
  for Fast Kernels}. In \bibinfo{booktitle}{{\em IFIP Working Conference on
  Domain-Specific Languages (DSL WC)}} {\em (\bibinfo{series}{Lecture Notes in
  Computer Science (LNCS)})}, Vol.~\bibinfo{volume}{5658}.
  \bibinfo{publisher}{Springer}, \bibinfo{pages}{385--410}.
\newblock


\bibitem[\protect\citeauthoryear{Frison, Kouzoupis, Zanelli, and Diehl}{Frison
  et~al\mbox{.}}{2017}]%
        {Frison:17}
\bibfield{author}{\bibinfo{person}{G. Frison}, \bibinfo{person}{D. Kouzoupis},
  \bibinfo{person}{A. Zanelli}, {and} \bibinfo{person}{M. Diehl}.}
  \bibinfo{year}{2017}\natexlab{}.
\newblock \showarticletitle{{BLASFEO:} Basic linear algebra subroutines for
  embedded optimization}.
\newblock \bibinfo{journal}{{\em CoRR\/}}  \bibinfo{volume}{abs/1704.02457}
  (\bibinfo{year}{2017}).
\newblock
\showURL{%
\url{http://arxiv.org/abs/1704.02457}}


\bibitem[\protect\citeauthoryear{Grosser, Groesslinger, and Lengauer}{Grosser
  et~al\mbox{.}}{2012}]%
        {Grosser:12}
\bibfield{author}{\bibinfo{person}{T. Grosser}, \bibinfo{person}{A.
  Groesslinger}, {and} \bibinfo{person}{C. Lengauer}.}
  \bibinfo{year}{2012}\natexlab{}.
\newblock \showarticletitle{Polly --- Performing Polyhedral Optimizations On A
  Low-Level Intermediate Representation}.
\newblock \bibinfo{journal}{{\em Parallel Processing Letters\/}}
  \bibinfo{volume}{22}, \bibinfo{number}{04} (\bibinfo{year}{2012}),
  \bibinfo{pages}{1250010}.
\newblock


\bibitem[\protect\citeauthoryear{Group}{Group}{2017}]%
        {Llvm}
\bibfield{author}{\bibinfo{person}{LLVM~Developer Group}.}
  \bibinfo{year}{2017}\natexlab{}.
\newblock \bibinfo{title}{The {LLVM} Compiler Infrastracture}.
\newblock \bibinfo{howpublished}{\url{http://llvm.org/}}.
\newblock


\bibitem[\protect\citeauthoryear{Guennebaud, Jacob, et~al\mbox{.}}{Guennebaud
  et~al\mbox{.}}{2017}]%
        {Eigen}
\bibfield{author}{\bibinfo{person}{G. Guennebaud}, \bibinfo{person}{B. Jacob},
  {et~al\mbox{.}}} \bibinfo{year}{2017}\natexlab{}.
\newblock \bibinfo{title}{Eigen}.
\newblock \bibinfo{howpublished}{\url{http://eigen.tuxfamily.org}}.
\newblock


\bibitem[\protect\citeauthoryear{Gunnels, Gustavson, Henry, and van~de
  Geijn}{Gunnels et~al\mbox{.}}{2001}]%
        {Gunnels:01b}
\bibfield{author}{\bibinfo{person}{J.~A. Gunnels}, \bibinfo{person}{F.~G.
  Gustavson}, \bibinfo{person}{G. Henry}, {and} \bibinfo{person}{R.~A. van~de
  Geijn}.} \bibinfo{year}{2001}\natexlab{}.
\newblock \showarticletitle{{FLAME}: Formal Linear Algebra Methods
  Environment.}
\newblock \bibinfo{journal}{{\em ACM Transactions on Mathematical Software
  (TOMS)\/}} \bibinfo{volume}{27}, \bibinfo{number}{4} (\bibinfo{year}{2001}),
  \bibinfo{pages}{422--455}.
\newblock


\bibitem[\protect\citeauthoryear{Guo, Bikshandi, Fraguela, Garzaran, and
  Padua}{Guo et~al\mbox{.}}{2008}]%
        {Guo:08}
\bibfield{author}{\bibinfo{person}{J. Guo}, \bibinfo{person}{G. Bikshandi},
  \bibinfo{person}{B.~B. Fraguela}, \bibinfo{person}{M.~J. Garzaran}, {and}
  \bibinfo{person}{D. Padua}.} \bibinfo{year}{2008}\natexlab{}.
\newblock \showarticletitle{Programming with tiles}. In
  \bibinfo{booktitle}{{\em Principles and Practice of Parallel Programming
  (PPoPP)}}. \bibinfo{pages}{111--122}.
\newblock


\bibitem[\protect\citeauthoryear{Heinecke, Henry, Hutchinson, and
  Pabst}{Heinecke et~al\mbox{.}}{2016}]%
        {Heinecke:16}
\bibfield{author}{\bibinfo{person}{A. Heinecke}, \bibinfo{person}{G. Henry},
  \bibinfo{person}{M. Hutchinson}, {and} \bibinfo{person}{H. Pabst}.}
  \bibinfo{year}{2016}\natexlab{}.
\newblock \showarticletitle{{LIBXSMM}: Accelerating Small Matrix
  Multiplications by Runtime Code Generation}. In \bibinfo{booktitle}{{\em High
  Performance Computing, Networking, Storage and Analysis (SC)}}.
  \bibinfo{pages}{1--11}.
\newblock


\bibitem[\protect\citeauthoryear{Intel}{Intel}{2017}]%
        {mkl}
\bibfield{author}{\bibinfo{person}{Intel}.} \bibinfo{year}{2017}\natexlab{}.
\newblock \bibinfo{title}{Intel Math Kernel Library ({MKL})}.
\newblock \bibinfo{howpublished}{\url{software.intel.com/en-us/intel-mkl}}.
\newblock


\bibitem[\protect\citeauthoryear{Jonsson and K{\aa}gstr\"{o}m}{Jonsson and
  K{\aa}gstr\"{o}m}{2002}]%
        {Jonsson:02}
\bibfield{author}{\bibinfo{person}{I. Jonsson} {and} \bibinfo{person}{B.
  K{\aa}gstr\"{o}m}.} \bibinfo{year}{2002}\natexlab{}.
\newblock \showarticletitle{Recursive Blocked Algorithms for Solving Triangular
  Systems --- {Part I}: One-sided and Coupled {Sylvester-type} Matrix
  Equations}.
\newblock \bibinfo{journal}{{\em ACM Transactions on Mathematical Software
  (TOMS)\/}} \bibinfo{volume}{28}, \bibinfo{number}{4} (\bibinfo{year}{2002}),
  \bibinfo{pages}{392--415}.
\newblock


\bibitem[\protect\citeauthoryear{Kong, Veras, Stock, Franchetti, Pouchet, and
  Sadayappan}{Kong et~al\mbox{.}}{2013}]%
        {Kong:13}
\bibfield{author}{\bibinfo{person}{M. Kong}, \bibinfo{person}{R. Veras},
  \bibinfo{person}{K. Stock}, \bibinfo{person}{F. Franchetti},
  \bibinfo{person}{L.~N. Pouchet}, {and} \bibinfo{person}{P. Sadayappan}.}
  \bibinfo{year}{2013}\natexlab{}.
\newblock \showarticletitle{When polyhedral transformations meet {SIMD} code
  generation}. In \bibinfo{booktitle}{{\em Programming Language Design and
  Implementation (PLDI)}}. \bibinfo{pages}{127--138}.
\newblock


\bibitem[\protect\citeauthoryear{Kyrtatas, Spampinato, and
  P{\"u}schel}{Kyrtatas et~al\mbox{.}}{2015}]%
        {Kyrtatas:15}
\bibfield{author}{\bibinfo{person}{N. Kyrtatas}, \bibinfo{person}{D.~G.
  Spampinato}, {and} \bibinfo{person}{M. P{\"u}schel}.}
  \bibinfo{year}{2015}\natexlab{}.
\newblock \showarticletitle{A Basic Linear Algebra Compiler for Embedded
  Processors}. In \bibinfo{booktitle}{{\em Design, Automation and Test in
  Europe (DATE)}}. \bibinfo{pages}{1054--1059}.
\newblock


\bibitem[\protect\citeauthoryear{Larsen and Amarasinghe}{Larsen and
  Amarasinghe}{2000}]%
        {Larsen:00}
\bibfield{author}{\bibinfo{person}{S. Larsen} {and} \bibinfo{person}{S.
  Amarasinghe}.} \bibinfo{year}{2000}\natexlab{}.
\newblock \showarticletitle{Exploiting Superword Level Parallelism with
  Multimedia Instruction Sets}. In \bibinfo{booktitle}{{\em Programming
  Language Design and Implementation (PLDI)}}. \bibinfo{pages}{145--156}.
\newblock


\bibitem[\protect\citeauthoryear{Lattner and Adve}{Lattner and Adve}{2004}]%
        {Lattner:04}
\bibfield{author}{\bibinfo{person}{C. Lattner} {and} \bibinfo{person}{V.
  Adve}.} \bibinfo{year}{2004}\natexlab{}.
\newblock \showarticletitle{{LLVM}: A Compilation Framework for Lifelong
  Program Analysis and Transformation}. In \bibinfo{booktitle}{{\em Code
  Generation and Optimization (CGO)}}. \bibinfo{pages}{75--86}.
\newblock


\bibitem[\protect\citeauthoryear{Low, Igual, Smith, and Quintana-Orti}{Low
  et~al\mbox{.}}{2016}]%
        {Low:16}
\bibfield{author}{\bibinfo{person}{T.~M. Low}, \bibinfo{person}{F.~D. Igual},
  \bibinfo{person}{T.~M. Smith}, {and} \bibinfo{person}{E.~S. Quintana-Orti}.}
  \bibinfo{year}{2016}\natexlab{}.
\newblock \showarticletitle{Analytical Modeling Is Enough for High-Performance
  {BLIS}}.
\newblock \bibinfo{journal}{{\em ACM Transactions on Mathematical Software
  (TOMS)\/}} \bibinfo{volume}{43}, \bibinfo{number}{2} (\bibinfo{year}{2016}),
  \bibinfo{pages}{12:1--12:18}.
\newblock


\bibitem[\protect\citeauthoryear{Marker, Poulson, Batory, and van~de
  Geijn}{Marker et~al\mbox{.}}{2013}]%
        {Marker:12}
\bibfield{author}{\bibinfo{person}{B. Marker}, \bibinfo{person}{J. Poulson},
  \bibinfo{person}{D. Batory}, {and} \bibinfo{person}{R. van~de Geijn}.}
  \bibinfo{year}{2013}\natexlab{}.
\newblock \showarticletitle{Designing Linear Algebra Algorithms by
  Transformation: Mechanizing the Expert Developer}.
\newblock In \bibinfo{booktitle}{{\em High Performance Computing for
  Computational Science (VECPAR 2012)}}. \bibinfo{series}{Lecture Notes in
  Computer Science (LNCS)}, Vol.~\bibinfo{volume}{7851}.
  \bibinfo{publisher}{Springer}, \bibinfo{pages}{362--378}.
\newblock


\bibitem[\protect\citeauthoryear{Nikolic, ElSeidy, and Koch}{Nikolic
  et~al\mbox{.}}{2014}]%
        {Nikolic:14}
\bibfield{author}{\bibinfo{person}{Milos Nikolic}, \bibinfo{person}{Mohammed
  ElSeidy}, {and} \bibinfo{person}{Christoph Koch}.}
  \bibinfo{year}{2014}\natexlab{}.
\newblock \showarticletitle{{LINVIEW}: Incremental View Maintenance for Complex
  Analytical Queries}. In \bibinfo{booktitle}{{\em Management of Data
  (SIGMOD)}}. \bibinfo{pages}{253--264}.
\newblock


\bibitem[\protect\citeauthoryear{Nuzman, Dyshel, Rohou, Rosen, Williams, Yuste,
  Cohen, and Zaks}{Nuzman et~al\mbox{.}}{2011}]%
        {Nuzman:11}
\bibfield{author}{\bibinfo{person}{D. Nuzman}, \bibinfo{person}{S. Dyshel},
  \bibinfo{person}{E. Rohou}, \bibinfo{person}{I. Rosen}, \bibinfo{person}{K.
  Williams}, \bibinfo{person}{D. Yuste}, \bibinfo{person}{A. Cohen}, {and}
  \bibinfo{person}{A. Zaks}.} \bibinfo{year}{2011}\natexlab{}.
\newblock \showarticletitle{Vapor {SIMD}: Auto-vectorize once, run everywhere}.
  In \bibinfo{booktitle}{{\em International Symposium on Code Generation and
  Optimization (CGO)}}. \bibinfo{pages}{151--160}.
\newblock


\bibitem[\protect\citeauthoryear{Nuzman, Rosen, and Zaks}{Nuzman
  et~al\mbox{.}}{2006}]%
        {Nuzman:06}
\bibfield{author}{\bibinfo{person}{D. Nuzman}, \bibinfo{person}{I. Rosen},
  {and} \bibinfo{person}{A. Zaks}.} \bibinfo{year}{2006}\natexlab{}.
\newblock \showarticletitle{Auto-vectorization of interleaved data for {SIMD}}.
  In \bibinfo{booktitle}{{\em Programming Language Design and Implementation
  (PLDI)}}. \bibinfo{pages}{132--143}.
\newblock


\bibitem[\protect\citeauthoryear{Peise and Bientinesi}{Peise and
  Bientinesi}{2016}]%
        {Peise:16}
\bibfield{author}{\bibinfo{person}{E. Peise} {and} \bibinfo{person}{P.
  Bientinesi}.} \bibinfo{year}{2016}\natexlab{}.
\newblock \showarticletitle{Recursive Algorithms for Dense Linear Algebra: The
  {ReLAPACK} Collection}.
\newblock  (\bibinfo{year}{2016}).
\newblock
\newblock
\shownote{Available at: http://arxiv.org/abs/1207.2169.}


\bibitem[\protect\citeauthoryear{Pharr and Mark}{Pharr and Mark}{2012}]%
        {Pharr:12}
\bibfield{author}{\bibinfo{person}{M. Pharr} {and} \bibinfo{person}{W.~R.
  Mark}.} \bibinfo{year}{2012}\natexlab{}.
\newblock \showarticletitle{ispc: A {SPMD} compiler for high-performance {CPU}
  programming}. In \bibinfo{booktitle}{{\em Innovative Parallel Computing
  (InPar)}}. \bibinfo{pages}{1--13}.
\newblock


\bibitem[\protect\citeauthoryear{P{\"u}schel, Franchetti, and
  Voronenko}{P{\"u}schel et~al\mbox{.}}{2011}]%
        {Pueschel:11}
\bibfield{author}{\bibinfo{person}{M. P{\"u}schel}, \bibinfo{person}{F.
  Franchetti}, {and} \bibinfo{person}{Y. Voronenko}.}
  \bibinfo{year}{2011}\natexlab{}.
\newblock \bibinfo{booktitle}{{\em Encyclopedia of Parallel Computing}}.
\newblock \bibinfo{publisher}{Springer}, Chapter Spiral.
\newblock


\bibitem[\protect\citeauthoryear{P{\"u}schel, Moura, Johnson, Padua, Veloso,
  Singer, Xiong, Franchetti, Gacic, Voronenko, Chen, Johnson, and
  Rizzolo}{P{\"u}schel et~al\mbox{.}}{2005}]%
        {Pueschel:05}
\bibfield{author}{\bibinfo{person}{M. P{\"u}schel}, \bibinfo{person}{J.~M.~F.
  Moura}, \bibinfo{person}{J. Johnson}, \bibinfo{person}{D. Padua},
  \bibinfo{person}{M. Veloso}, \bibinfo{person}{B. Singer}, \bibinfo{person}{J.
  Xiong}, \bibinfo{person}{F. Franchetti}, \bibinfo{person}{A. Gacic},
  \bibinfo{person}{Y. Voronenko}, \bibinfo{person}{K. Chen},
  \bibinfo{person}{R.~W. Johnson}, {and} \bibinfo{person}{N. Rizzolo}.}
  \bibinfo{year}{2005}\natexlab{}.
\newblock \showarticletitle{{SPIRAL}: Code Generation for {DSP} Transforms}.
\newblock \bibinfo{journal}{{\it Proc. IEEE}} \bibinfo{volume}{93},
  \bibinfo{number}{2} (\bibinfo{year}{2005}), \bibinfo{pages}{232--275}.
\newblock


\bibitem[\protect\citeauthoryear{Rapaport, Zaks, and Ben-Asher}{Rapaport
  et~al\mbox{.}}{2015}]%
        {Rapaport:15}
\bibfield{author}{\bibinfo{person}{G. Rapaport}, \bibinfo{person}{A. Zaks},
  {and} \bibinfo{person}{Y. Ben-Asher}.} \bibinfo{year}{2015}\natexlab{}.
\newblock \showarticletitle{Streamlining Whole Function Vectorization in C
  Using Higher Order Vector Semantics}. In \bibinfo{booktitle}{{\em Parallel
  and Distributed Processing Symposium Workshop (IPDPSW)}}.
  \bibinfo{pages}{718--727}.
\newblock


\bibitem[\protect\citeauthoryear{Rasmussen and Williams}{Rasmussen and
  Williams}{2005}]%
        {Rasmussen:05}
\bibfield{author}{\bibinfo{person}{C.~E. Rasmussen} {and}
  \bibinfo{person}{C.~K.~I. Williams}.} \bibinfo{year}{2005}\natexlab{}.
\newblock \bibinfo{booktitle}{{\em Gaussian Processes for Machine Learning
  (Adaptive Computation and Machine Learning)}}.
\newblock \bibinfo{publisher}{The MIT Press}.
\newblock
\showISBNx{026218253X}


\bibitem[\protect\citeauthoryear{Scharf}{Scharf}{1991}]%
        {Scharf:91}
\bibfield{author}{\bibinfo{person}{L.~L. Scharf}.}
  \bibinfo{year}{1991}\natexlab{}.
\newblock \bibinfo{booktitle}{{\em Probability, Statistical Signal Processing,
  Detection, Estimation, and Time Series Analysis.}}
\newblock \bibinfo{publisher}{Addison-Wesley}.
\newblock


\bibitem[\protect\citeauthoryear{Siek, Karlin, and Jessup}{Siek
  et~al\mbox{.}}{2008}]%
        {Siek:08}
\bibfield{author}{\bibinfo{person}{J.G. Siek}, \bibinfo{person}{I. Karlin},
  {and} \bibinfo{person}{E.R. Jessup}.} \bibinfo{year}{2008}\natexlab{}.
\newblock \showarticletitle{Build to order linear algebra kernels}. In
  \bibinfo{booktitle}{{\em International Parallel \& Distributed Processing
  Symposium (IPDPS)}}. \bibinfo{pages}{1--8}.
\newblock


\bibitem[\protect\citeauthoryear{Spampinato and P{\"u}schel}{Spampinato and
  P{\"u}schel}{2014}]%
        {Spampinato:14}
\bibfield{author}{\bibinfo{person}{D.~G. Spampinato} {and} \bibinfo{person}{M.
  P{\"u}schel}.} \bibinfo{year}{2014}\natexlab{}.
\newblock \showarticletitle{A Basic Linear Algebra Compiler}. In
  \bibinfo{booktitle}{{\em International Symposium on Code Generation and
  Optimization (CGO)}}. \bibinfo{pages}{23--32}.
\newblock


\bibitem[\protect\citeauthoryear{Spampinato and P{\"u}schel}{Spampinato and
  P{\"u}schel}{2016}]%
        {Spampinato:16}
\bibfield{author}{\bibinfo{person}{D.~G. Spampinato} {and} \bibinfo{person}{M.
  P{\"u}schel}.} \bibinfo{year}{2016}\natexlab{}.
\newblock \showarticletitle{A Basic Linear Algebra Compiler for Structured
  Matrices}. In \bibinfo{booktitle}{{\em International Symposium on Code
  Generation and Optimization (CGO)}}. \bibinfo{pages}{117--127}.
\newblock


\bibitem[\protect\citeauthoryear{{The MathWorks, Inc.}}{{The MathWorks,
  Inc.}}{2017}]%
        {MatlabCoder}
\bibfield{author}{\bibinfo{person}{{The MathWorks, Inc.}}}
  \bibinfo{year}{2017}\natexlab{}.
\newblock \bibinfo{title}{{MATLAB} Coder}.
\newblock
  \bibinfo{howpublished}{\url{https://www.mathworks.com/products/matlab-coder.html}}.
\newblock


\bibitem[\protect\citeauthoryear{Van~Zee and van~de Geijn}{Van~Zee and van~de
  Geijn}{2015}]%
        {VanZee:15}
\bibfield{author}{\bibinfo{person}{F.~G. Van~Zee} {and} \bibinfo{person}{R.~A.
  van~de Geijn}.} \bibinfo{year}{2015}\natexlab{}.
\newblock \showarticletitle{BLIS: A Framework for Rapidly Instantiating BLAS
  Functionality}.
\newblock \bibinfo{journal}{{\em ACM Transactions on Mathematical Software
  (TOMS)\/}} \bibinfo{volume}{41}, \bibinfo{number}{3}, Article
  \bibinfo{articleno}{14} (\bibinfo{year}{2015}), \bibinfo{numpages}{33}~pages.
\newblock


\bibitem[\protect\citeauthoryear{Walter, Koch, et~al\mbox{.}}{Walter
  et~al\mbox{.}}{2012}]%
        {uBlas}
\bibfield{author}{\bibinfo{person}{J. Walter}, \bibinfo{person}{M. Koch},
  {et~al\mbox{.}}} \bibinfo{year}{2012}\natexlab{}.
\newblock \bibinfo{title}{{uBLAS}}.
\newblock \bibinfo{howpublished}{\url{www.boost.org/libs/numeric/ublas}}.
\newblock


\bibitem[\protect\citeauthoryear{Whaley and Dongarra}{Whaley and
  Dongarra}{1998}]%
        {Whaley:98}
\bibfield{author}{\bibinfo{person}{R.~C. Whaley} {and} \bibinfo{person}{J.~J.
  Dongarra}.} \bibinfo{year}{1998}\natexlab{}.
\newblock \showarticletitle{Automatically tuned linear algebra software}. In
  \bibinfo{booktitle}{{\em Supercomputing (SC)}}. \bibinfo{pages}{1--27}.
\newblock


\bibitem[\protect\citeauthoryear{Zhang}{Zhang}{2017}]%
        {OpenBLAS}
\bibfield{author}{\bibinfo{person}{X. Zhang}.} \bibinfo{year}{2017}\natexlab{}.
\newblock \bibinfo{title}{The {OpenBLAS} library}.
\newblock \bibinfo{howpublished}{\url{http://www.openblas.net/}}.
\newblock


\end{thebibliography}

\end{document}